\documentclass[traditabstract]{aa}
\usepackage{epsfig}    
\usepackage{lscape}
\usepackage{natbib}
\bibpunct{(}{)}{;}{a}{}{,}    
\usepackage{graphics}
\usepackage{graphicx,graphics}
\usepackage{color} 
\usepackage{times}
\usepackage{amsmath}
\usepackage{amssymb}
\begin{document}

\title {Metal-THINGS: The Milky Way twin candidate NGC~3521} 

 \author {
          L.~S.~Pilyugin\inst{\ref{ITPA},\ref{MAO}}         \and
          M.~A.~Lara-L\'{o}pez\inst{\ref{UCM}}             \and
          G.~Tautvai\v{s}ien\.{e}\inst{\ref{ITPA}}         \and
          I. A. Zinchenko\inst{\ref{LMU},\ref{MAO}}       \and
          L. E. Gardu\~{n}o\inst{\ref{INAOE}}            \and
          M. E. De Rossi\inst{\ref{UB},\ref{IAFE}}         \and
          J.~Zaragoza-Cardiel \inst{\ref{CEFCA}}           \and
          S.~Dib\inst{\ref{MPI}}                         \and
          G. Val\'{e}\inst{\ref{UCM}}
 }
 
\institute{
   Institute of Theoretical Physics and Astronomy, Vilnius University, Sauletekio av. 3, 10257, Vilnius, Lithuania \label{ITPA}             \and
   Main Astronomical Observatory, National Academy of Sciences of Ukraine, 27 Akademika Zabolotnoho St, 03680, Kiev, Ukraine \label{MAO}    \and
   Departamento de F\'{i}sica de la Tierra y Astrof\'{ı}sica, Instituto de F\'{ı}sica de Part\'{ı}culas y del Cosmos, IPARCOS. Universidad
      Complutense de Madrid (UCM), E-28040, Madrid, Spain \label{UCM}                                                                        \and
   Faculty of Physics, Ludwig-Maximilians-Universit{\"a}t, Scheinerstr.1, 81679 Munich, Germany\label{LMU}                                   \and
   Instituto Nacional de Astrof\'{ı}sica,  \'{O}ptica y Electr\'{o}nica (INAOE), Luis E. Erro 1, Tonantzintla, Puebla, C.P. 72840,
      M\'{e}xico\label{INAOE}                                                                                                                \and
   Universidad de Buenos Aires, Facultad de Ciencias Exactas y Naturales y Ciclo B\'{a}sico Com\'{u}n. Buenos Aires, Argentina\label{UB}     \and
   CONICET-Universidad de Buenos Aires, Instituto de Astronom\'{i}a y F\'{i}sica del Espacio (IAFE). Buenos Aires, Argentina\label{IAFE}     \and
   Centro de Estudios de Física del Cosmos de Aragón (CEFCA), Plaza San Juan 1, 44001 Teruel, Spain\label{CEFCA}                             \and
   Max Planck Institute for Astronomy, K\"{o}nigstuhl 17, 69117, Heidelberg, Germany\label{MPI}
       }

\abstract{
The 3D spectrophotometry measurements of the galaxy NGC~3521, a structural Milky Way analogue (sMWA), were carried out within the Metal-THINGS project.
We found that the oxygen abundance in the inner part of  NGC~3521 is at a nearly constant level and the O/H gradient is negative at larger radii.
The change in the nitrogen abundance with radius is similar to that for oxygen with the break in the N/H distribution at a smaller radius than the O/H distribution break,
but the difference between the break radii is within the uncertainties of these values.
The radial distributions of the oxygen abundance, the gas mass fraction, and the effective oxygen yield in NGC~3521 are compared to that of the Milky Way (MW), with the  aim
of examining the similarity (or disagreement) in their chemical evolutions.
The oxygen abundances of two H\,{\sc ii} regions closest to the centre of the MW (at a radii of 4-5 kpc) are close to the binned oxygen abundances in NGC~3521 at the
same galactocentric distances; an accurate value of the central oxygen abundance in the MW cannot be established because of the lack of the measurements near the centre. 
The oxygen abundances in the outer part of the MW are lower than those in the outer part of NGC~3521. The gas mass fraction in the outer part of the MW is higher than in NGC~3521. 
The obtained values of the effective oxygen yield, $Y_{eff}$, in  NGC~3521 are close to the empirical estimation of the oxygen yield, $Y_{O}$.
This suggests that mass exchange with the surroundings plays little to no role in the current chemical evolution of NGC3521.
The values of the $Y_{eff}$ in the MW were determined using two variants of the  radial distribution of the gas mass surface density. The values of the $Y_{eff}$ in the MW obtained with
the first  distribution are also close to  $Y_{O}$, as  in NGC~3521. The $Y_{eff}$ in the MW obtained with the second distribution are below $Y_{O}$ at radii between $\sim$6 and $\sim$10.4 kpc.
This suggests that the mass exchange with the surroundings can play a significant role in the chemical evolution of this part of the MW, in contrast to that in  NGC~3521. 
To draw a solid conclusion about the role of  mass exchange  with the surroundings in the chemical evolution of the MW it is essential to determine which of these distributions provides
a more adequate description of the gas distribution in the MW.
}


\titlerunning{Metal-THINGS: NGC~3521 versus the Milky Way}
\authorrunning{Pilyugin et al.}
\maketitle

\section{Introduction}

The examination of the oxygen abundance distributions in structural Milky Way analogues (sMWAs) is important. Since the oxygen abundance is an indicator of the evolution of a galaxy,
the comparison of the oxygen abundance distributions between the Milky Way (MW) and sMWAs, coupled with the comparison of other relevant characteristics (e.g. gas mass fraction),
can point out  whether the chemical evolution of the MW is similar to (or differs from) the chemical evolutions of the sMWAs. This can clarify whether the MW is really a typical spiral
galaxy, and in what ways it differs from typical spiral galaxies if this is not the case. This is important for the theory of formation and evolution of galaxies. The search for
Milky Way-like galaxies has been the subject of many investigations \citep[e.g.][]{deVaucouleurs1978,Hammer2007,Mutch2011,Licquia2015b,Licquia2016b,McGaugh2016,FraserMcKelvie2019,
Boardman2020a,Fielder2021,Pilyugin2024}. It was noted by \citet{Boardman2020a} that there is no single and commonly accepted definition of a Milky Way-like galaxy; rather, the definition
can change depending on the goals of a particular study. Different characteristics of the MW (morphological type $T$, isophotal diameter, effective diameter, absolute $B$ magnitude,
mean colour index ($B - V$), presence of spiral arms, presence of a bar, stellar mass,  bulge-to-total ratio, rotation velocity) have been used when comparing the MW to other galaxies. 
Galaxies may be identified as being Milky Way-like on the basis of having similar qualitative characteristics to the MW or on the basis of their position relative to the MW in a given
parameter space. The selection of Milky Way-like galaxies using only two characteristics of the MW as the selection criteria is based on the Copernican assumption that the MW is not
extraordinary amongst galaxies; that is, any property of the MW is similar to that of the Milky Way-like galaxies selected \citep{Mutch2011,Licquia2015b,Boardman2020a}.

It was suggested to distinguish between structural Milky Way analogues and evolutionary Milky Way analogues \citep{Pilyugin2024}. The characteristics of a galaxy can be conditionally
divided into two types. The parameters of the first type (e.g. morphology, luminosity, stellar mass, rotation velocity) describe the structure and global characteristics of a galaxy
at the present-day epoch, and can be called structural parameters. The parameters of the second type are related to the  evolution of a galaxy. The oxygen abundance at a given radius
of a galaxy is defined by the evolution of this region (fraction of gas converted into stars, i.e. astration level, and matter exchange with the surroundings). Thus, the oxygen abundance
can be considered as an indicator of a galaxy's evolution and can be called  an evolutionary parameter. A galaxy located close to the MW in the field(s) of the first type of parameters
should be referred to as a structural Milky Way analogue (sMWA). A galaxy located close to the MW in the field of the second type of parameters should be referred to as an evolutionary
Milky Way analogue (eMWA). If a galaxy is simultaneously an sMWA and an eMWA, then such galaxy should be considered a Milky Way twin.

sMWAs were selected and examined in the papers cited above. It was found that the galaxy NGC~3521 has structural parameters (stellar mass, rotation velocity, optical radius) that are
very similar to those of the MW; these quantities are identical within the uncertainties, which means that NGC~3521 is a near structural analogue to the MW \citep{McGaugh2016}. It is
interesting to note that the mass of the supermassive black hole at the centre of the Milky Way, $M_{BH}$ = 4.15$\pm$0.01$\times$10$^{6}M_{\sun}$ or log($M_{BH}/M_{\sun}$) = 6.618,
derived using the star orbit around the black hole  \citep{Gravity2019} is also close to the supermassive black hole mass in  NGC~3521, log($M_{BH}/M_{\sun}$) = 6.85$\pm$0.58, estimated
using an established relationship between the logarithmic spiral arm pitch angle and the mass of the central supermassive black hole in their host galaxies \citep{Davis2014}.

The spectra of ten H\,{\sc ii} regions in NGC~3521 were measured by \citet{Zaritsky1994}, three H\,{\sc ii} regions by \citet{Bresolin1999}, and one H\,{\sc ii} region by
\citet{Lopez2019}.  A mandatory requirement for comparing the oxygen abundances in the MW and NGC~3521 is that the abundances must be determined using the same metallicity scale.
The oxygen abundances in the H\,{\sc ii} regions of the MW are determined through the direct $T_{e}$ method. Therefore, we used the R calibration from \citet{Pilyugin2016} in the
determinations of the oxygen abundances in  H\,{\sc ii} regions of NGC~3521 since the R calibration provides abundances compatible with the metallicity scale defined by the
$T_{e}$-based abundances. Unfortunately, \citet{Zaritsky1994} did not measure the fluxes of the nitrogen emission line [N\,{\sc ii}]$\lambda$6584 required for the abundance determination
through the R calibration. Thus, the R calibration-based abundances can be estimated only in the  H\,{\sc ii} regions of NGC~3521 measured by \citet{Bresolin1999} and \citet{Lopez2019}.
\citet{Grasha2022} identified 92  H\,{\sc ii} regions in NGC~3521 using  data from the TYPHOON programme, an integral field spectroscopic survey of 44 nearby, large angular-size galaxies
observed with the 2.5~m du Pont Telescope at the Las Campanas Observatory, Chile. \citet{Grasha2022} measured the emission-line fluxes of the  H\,{\sc ii} regions and determined their
oxygen abundances using the N2O2 calibration from \citet{Kewley2002}. They measured the radial oxygen abundance gradient across NGC~3521. However, the values they used for the geometric
parameters  (e.g. the optical radius and the position angle of the major axis) of NGC~3521 are inaccurate. Their adopted value of the optical radius $R_{25}$ is less than a half of that
reported in the RC3 catalogue \citep{RC3}. Their adopted value of the position angle of the major axis is PA=107$\degr$, while PA is $\sim$163$\degr$, north-eastwards, in RC3 and other
catalogues. The value  of PA $\sim$341$\degr$ (or PA $\sim$161$\degr$, north-eastwards) was derived  from the analysis of the velocity fields of the ionised hydrogen, neutral hydrogen,
and CO (2–1) emission \citep{Daigle2006,deBlok2008,Lang2020}. Taking into account that NGC~3521 is a galaxy with a high inclination, the use of the correct value of the position angle of
the major axis is crucial  in the determinations of the galactocentric distances of  H\,{\sc ii} regions. Therefore, the value of the oxygen abundance gradient obtained in \citet{Grasha2022}
is unreliable. This hinders the ability to determine the similarities or discrepancies in the evolutionary paths of NGC 3521 and the MW, making it difficult to conclude whether NGC 3521 is
a true evolutionary analogue of the Milky Way.

Three pointings of  NGC~3521 were observed with integral field spectroscopy as part of the Metal-THINGS survey \citep{LaraLopez2021, LaraLopez2023}. The blue and red spectra were observed,
and the emission lines necessary for  abundance determinations through the R calibration from \citet{Pilyugin2016} were measured. This allowed the derivation of the radial abundance
distribution across  NGC~3521 on a metallicity scale that is compatible with the $T_{e}$-based oxygen abundances in the MW. By comparing the radial distributions of the oxygen abundances,
gas mass fractions, and effective oxygen yields in NGC 3521 and the MW, we can identify similarities or discrepancies in their chemical evolutions. This is the goal of the current study.

This paper is organised in the following way. The observations and data reduction of  NGC~3521 are described in Sect. 2. In Sect. 3 the determinations of the orientation of NGC~3521 in
space, the rotation curve, and the radial abundance distribution are reported. The comparison between  NGC~3521 and the Milky Way is given in Sect. 4, and Sect. 5 provides a brief summary.

\section{Observations and data reduction}

\begin{figure}
\resizebox{0.8\hsize}{!}{\includegraphics[angle=000]{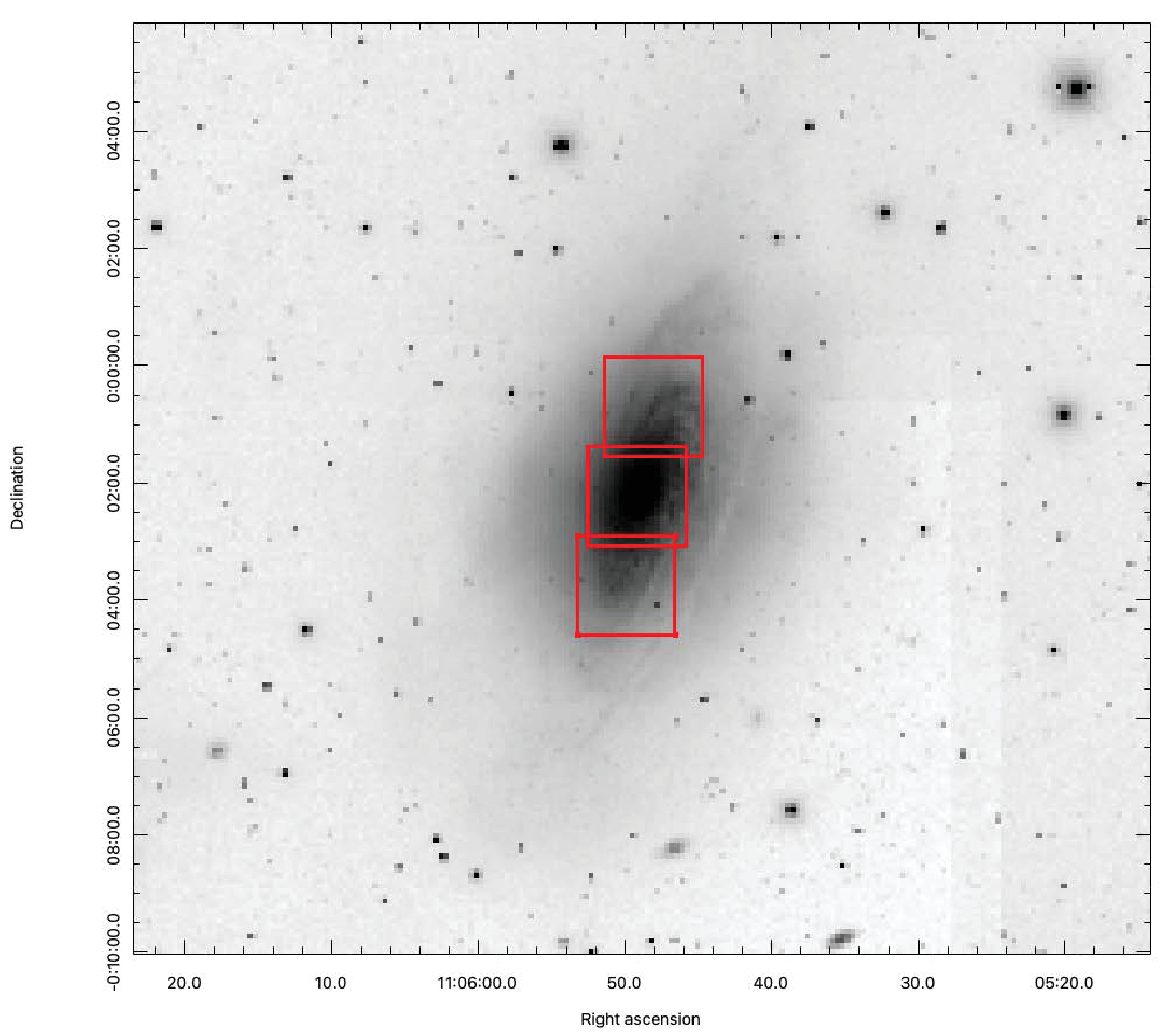}}
\caption{
Observed GCMS pointings superimposed on a NGC~3521 image from the Cerro Tololo Inter-American Observatory (CTIO), 1.5m telescope. 
} 
\label{figure:pointings}
\end{figure}

The observations of NGC~3521 were carried out as part of the Metal-THINGS survey using the 2.7 m Harlan Schmidt telescope, McDonald Observatory, Texas, in March 2022. The Integral Field
Unit (IFU) George and Cynthia Mitchell Spectrograph  \citep[GCMS,][]{Hill2008}  was used in the blue and red setup with the low-resolution grating VP1 with a resolution of 5.3 \AA.
The average seeing during the observations was 1.7 arsec, and a total of three pointings were observed, as shown in Fig.~\ref{figure:pointings}. 

The GCMS is a square array of 100 $\times$ 102 arcsec, with a spatial sampling of 4.2 arcsec, and a 0.3 filling factor. Every pointing is observed with three dither positions to ensure
a 90\% surface coverage. The IFU consists of 246 fibres arranged in a fixed pattern. Each fibre has a 4.2 arcsec diameter, and hence that is the approximate spatial resolution. 
At the adopted distance of 10.7 Mpc to this galaxy (see below), the 4.2 arcsec fibre corresponds to 217.3 parsec. Due to the extended nature of this galaxy, off-source sky exposures were
taken for sky subtraction during the data reduction process. Every pointing was observed for 900 seconds per dither, followed by a sky exposure, and the same process was repeated until 
a total of 45 minutes per dither was reached. 

The data reduction was performed as described in \citet{LaraLopez2021, LaraLopez2023}. The basic data reduction (bias subtraction, flat frame correction, and wavelength calibration)
was performed using P3D.\footnote{https://p3d.sourceforge.io} The rest of the data reduction, including sky subtraction, flux calibration, combination of dithers, and mosaic generation,
was performed using our own routines in Python. Since we used several individual pointings for the same galaxy, astrometry was applied to each pointing. First, each pointing was converted
into a collapsed data cube; then we identified several stars and used the same star positions from the Two Micron All-Sky Survey \citep[2MASS,][]{Skrutskie2006}. Next we applied the astrometry
to each pointing using our own routines in Python (for more details, see \citealt{Garduno2023}). Finally, we assembled all the individual pointings and built a mosaic for this galaxy. 

The stellar continuum of all flux-calibrated spectra was fitted using STARLIGHT (\citealt{CidFernandes2005, Asari2007}; for a detailed description of this procedure, see  \citealt{Zinchenko2016}).
In summary, to fit the continuum 45 simple stellar population (SSP) models from the evolutionary synthesis models of  \citet{BruzualCharlot2003} were used, with ages from 1 Myr up to 13 Gyr
and metallicities Z = 0:005, 0.02, and 0.05.  The fitted continuum was subtracted from the spectra, and the emission lines were measured using Gaussian line-profile fittings.  In this paper
the individual spectra of the fibres are used, which are independent of each other. This is in contrast to other IFU data that rely on image spaxels.

\section{Determination of the properties of  NGC~3521}

\subsection{Orientation in the space and rotation curve}

\begin{figure}
\resizebox{1.00\hsize}{!}{\includegraphics[angle=000]{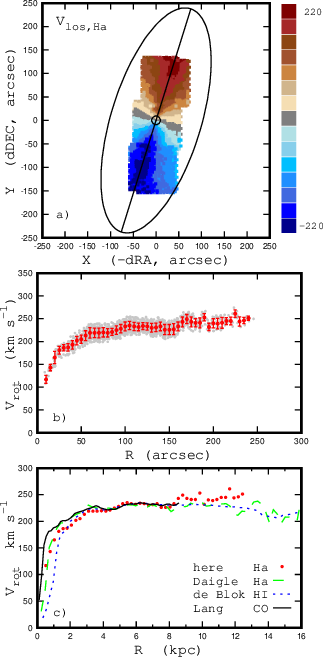}}
\caption{
Rotation curve of  NGC~3521.
{\em Panel} {\bf a:} Line-of-sight H$\alpha$ velocity (km s$^{-1}$) field in sky coordinates (arcsec). The values of the  $V_{los}$ are colour-coded.
The circle shows the kinematic centre of the galaxy; the line indicates the position of the major kinematic axis of the galaxy; the ellipse is the
optical radius. 
{\em Panel} {\bf b:} Rotation  curve of  NGC~3521. The grey points mark the individual fibres used in the determination of the final rotation curve.
The red circles denote the mean values of the rotation velocity in bins of 5 arcsec in the radius, and the bars show the scatter in rotation velocities 
about the mean value in the bins. 
{\em Panel} {\bf c:} Comparison of the obtained rotation curve with those from the literature. 
The red points show the rotation curve obtained here from the H$\alpha$ velocity field;
the green dashed line denotes the rotation curve determined by \citet{Daigle2006}  from the H$\alpha$ velocity field;
the  blue dotted line marks the rotation curve derived by  \citet{deBlok2008}  from the  H\,{\sc i} velocity field;
and the dark solid line is the rotation curve derived by \citet{Lang2020} from the  CO velocity field. 
}
\label{figure:rc}
\end{figure}

The geometrical parameters of the galaxy (such as the coordinates of its centre, the position angle of the major axis, the inclination angle, and the isophotal radius) are necessary
for determining the galactocentric distances of individual fibres and constructing radial distributions of various characteristics. The orientation of the galaxy in space can be derived
by the best fit to the observed line-of-sight velocity field. The measured wavelength of the observed H$\alpha$ emission line provides the line-of-sight velocity of each region (fibre),
denoted as $V_{\rm H\alpha}$. The observed line-of-sight velocities, $V_{los}$, are recorded on a set of fibre coordinates RA and DEC. Panel (a) of Fig.~\ref{figure:rc} shows the line-of-sight
velocity $V_{los}$ field across the image of NGC~3521 in sky coordinates (arcsec) where the measured $V_{\rm H\alpha}$ velocity is corrected for the obtained velocity of the centre of the
galaxy with respect to the Sun, the systemic velocity $V_{sys}$, that is  $V_{los}$ = $V_{\rm H\alpha}$ - $V_{sys}$.

The determination of the galaxy's geometrical  parameters  and the rotation curve from the observed velocity field is performed using standard method 
\citep[e.g.][]{Warner1973,Begeman1989,deBlok2008,Oh2018}.  
The parameters that define the observed velocity field of a galaxy with a symmetrically rotating disc  are as follows: \\
-- the coordinates  of the galaxy's rotation centre, RA$_{centre}$ and DEC$_{centre}$; \\
-- the velocity of the galaxy's centre relative to the Sun, the systemic velocity $V_{sys}$; \\ 
-- the circular velocity $V_{rot}$ at the distance $R$ from the galaxy's centre; \\ 
-- the position angle $PA$ of the major kinematic axis; \\
-- the kinematic inclination angle $i$, which is the angle between the normal to the plane of the galaxy rotation and the line of sight.

The parameters  RA$_{centre}$, DEC$_{centre}$, $V_{sys}$, $PA$, $i$, and the rotation curve  $V_{rot}(R)$ are derived by the best fit to the observed velocity field $V_{los}$(RA,DEC).  
We used the iterative procedure discussed in \citet{Pilyugin2019}. In brief, in the first step the values of RA$_{centre}$, DEC$_{centre}$, $PA$, $i$, and the rotation curve are determined
using all the available  line-of-sight velocity measurements. At each step, data points with large deviations (larger than 21 km s$^{-1}$, which is 3$\sigma$ for differences between the
measured H$\beta$ and H$\alpha$ velocities; \citealt{Pilyugin2019}) from the rotation curve obtained in the previous step are rejected, and new values of the parameters and of the
rotation curve  $V_{rot}(R)$ are derived. The iteration is stopped when the absolute value of the difference of RA$_{centre}$ (and  DEC$_{centre}$) obtained in successive steps is less than
0$\farcs$1, the difference of $PA$ (and $i$) is smaller than 0$\fdg$1, and the rotation curves agree within 1 km/s (at each radius).  

The following values of the geometrical parameters were derived: RA$_{centre}$ = 166$\fdg$451752, DEC$_{centre}$ = --0$\fdg$035979, $V_{sys}$ = 709.1 km/s, $PA$ = 342$\fdg$2, and $i$ = 66$\fdg$7.
The obtained position of the rotation centre of the galaxy and the position angle of the major kinematic axis are shown in panel (a) of Fig.~\ref{figure:rc}. The derived rotation curve of
NGC~3521 is shown  in panel (b) of Fig.~\ref{figure:rc}. The geometrical angles and the rotation curve derived from the stellar line-of-sight velocity field are close to those derived from
the line-of-sight velocity field of the ionised gas. 

We now compare the geometrical angles of  NGC~3521 obtained here with those from the literature. \citet{Daigle2006}  measured the H$\alpha$ line-of-sight velocity field in NGC~3521 based
on Fabry–Perot observations. From their analysis of the observed H$\alpha$ velocity field, they derived the kinematical angles and the rotation, finding   $PA$ = 342$\fdg$0$\pm$1$\fdg$1
and an inclination angle of $i$ = 66$\fdg$7$\pm$2$\degr$.

NGC~3521 was also studied as part of The H\,{\sc i} Nearby Galaxy Survey (THINGS) project, a high spectral ($\la$ 5.2 km/s) and spatial ($\sim$6$\arcsec$) resolution survey of  H\,{\sc i}
21 cm emission in 34 nearby galaxies obtained using the NRAO Very Large Array (VLA) \citep{Walter2008}. Using the velocity field of the H\,{\sc i} emission line, \citet{deBlok2008}
determined the position angle of the major kinematic axis, the inclination angle, and the rotation curve for NGC~3521, reporting $PA$ = 339$\fdg$8 and  $i$ = 72$\fdg$7. NGC~3521 is also
part of the sample of galaxies  of the Physics at High Angular resolution in Nearby Galaxies (PHANGS)--Atacama Large Millimeter/submillimeter Array (ALMA) project (PHANGS--ALMA;
\citealt{Leroy2021}). The CO (2–1) emission in this sample of galaxies is mapped at high ($\sim$1$\arcsec$) angular resolution. Using the CO data from PHANGS--ALMA,  \citet{Lang2020}
derived the kinematical angles and rotation curve for  NGC~3521, finding  $PA$ = 343$\fdg$0 and  $i$ = 68$\fdg$8. Thus, the geometrical angles obtained in this study are in excellent
agreement with those of  \citet{Daigle2006} for the velocity field of  ionised hydrogen. The differences in the angles based on the velocity fields of ionised hydrogen and molecular gas are
within $\sim$2$\degr$, while the difference in inclination angle between the ionised and atomic gas measurements is approximately $\sim$6$\degr$. The geometrical angles obtained here are used
to determine the galactocentric distances of individual fibres. 

Different distances to  NGC~3521 have been adopted in different studies: d = 9 Mpc in \citet{Daigle2006}, d = 10.7 Mpc in \citet{deBlok2008}, d = 11.2 Mpc in  \citet{Lang2020}.   
We adopted the intermediate value, the distance of d = 10.7 Mpc used in the THINGS project.  We also adopted an optical radius for NGC~3521 of $R_{25}$ = 4.16 arcmin or 249.6 arcsec from
the HyperLeda\footnote{http://leda.univ-lyon1.fr/} database \citep{Makarov2014}.  At this adopted distance the physical optical radius of  NGC~3521 is $R_{25}$ = 12.94 kpc. 
The chosen values of the distance and optical (isophotal) radius are used to convert the galactocentric distances of individual fibres from arcsec to   galactocentric distances in
a physical scale (kpc), and to the fractional galactocentric distances (normalised to the optical radius) in the galaxy plane.

In panel (c) of Fig.~\ref{figure:rc} we compare the rotation curve obtained here using the H$\alpha$ velocity field with those from previous studies determined using
the H$\alpha$ velocity field \citep{Daigle2006},  the  H\,{\sc i} velocity field \citep{deBlok2008}, and  the CO velocity field \citep{Lang2020}.  
Panel (c) of Fig.~\ref{figure:rc} reveals a significant discrepancy between the rotation curves derived from different velocity indicators  in  the central region of
the galaxy ($\la$ 2 kpc), while the two curves based on H$\alpha$ velocity fields are relatively close to each other.
Our rotation curve shows satisfactory agreement with the others in the range 3 kpc $\la$ R $\la$ 8.5 kpc, though it is slightly higher at larger radii.
The discrepancy at  large radii may be due to the fact that our measurements along the galaxy's major axis cover only $\sim$8.5 kpc (see panel (a) of Fig.~\ref{figure:rc}). 
Data points near the major axis provide the most reliable information for the determination of the rotation curve.

\subsection{Radial abundance distribution}

\begin{figure}
\resizebox{1.00\hsize}{!}{\includegraphics[angle=000]{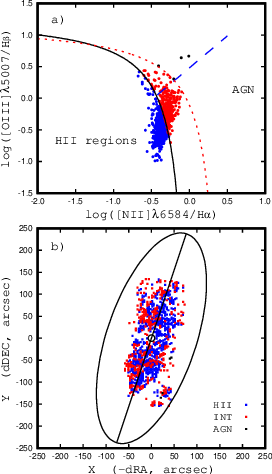}}
\caption{
  BPT types of the fibre spectra in  NGC~3521. 
  {\em Panel} {\bf a:}  BPT diagram for the individual fibres in  NGC~3521. The blue symbols denote the fibres with   H\,{\sc ii}-region-like spectra,
  the red symbols mark the fibres with  intermediate BPT-type spectra, and the black symbols are the fibres with  AGN-like spectra. 
  The solid and short-dashed curves mark the demarcation lines between AGNs and H\,{\sc ii} regions defined by \citet{Kauffmann2003} and \citet{Kewley2001}, respectively. 
  The long-dashed line is the dividing line between Seyfert galaxies and LINERs defined by \citet{CidFernandes2010}.
  {\em Panel} {\bf b:} Distribution of the fibres with different BPT-type spectra across the image of  NGC~3521. The BPT types of spectra are colour-coded 
  as in panel (a). The circle shows the kinematic centre of the galaxy, the line indicates the position of the major kinematic axis of the galaxy, and the ellipse corresponds to
  the optical radius of the galaxy. 
}
\label{figure:bpt}
\end{figure}

\begin{figure}
\resizebox{1.00\hsize}{!}{\includegraphics[angle=000]{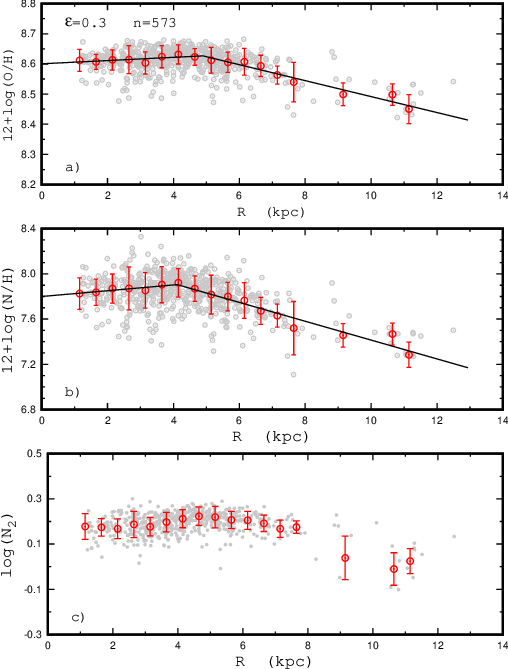}}
\caption{Abundance properties of NGC~3521 traced by the fibres where the line flux errors $\varepsilon <$ 0.3.
  {\em Panel} {\bf a:} Radial oxygen abundance distribution. The grey circles denote the $R$ calibration-based abundances for the individual fibres.
  The red circles denote the median values of the oxygen abundances in bins of 0.5 kpc in radius, and the bars show the scatter in O/H 
  around the median value of each bin. 
  The line shows the O/H -- $R$ relation for those data. 
  {\em Panel} {\bf b:} Same as panel (a), but for the nitrogen abundances. 
  {\em Panel} {\bf c:} Intensity of the emission nitrogen line N$_{2}$ as a function of radius. The notations are as in panel (a). 
}
\label{figure:r-oh-nh}
\end{figure}

\begin{figure}
\resizebox{1.00\hsize}{!}{\includegraphics[angle=000]{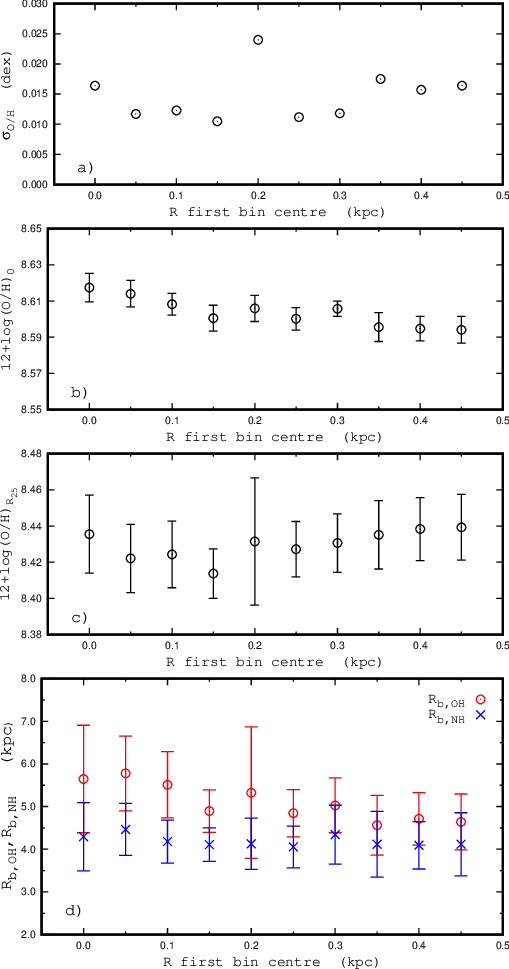}}
\caption{Dependence of the parameters of the derived oxygen abundance distribution in NGC~3521 on the positions of the centres of bins.
  {\em Panel} {\bf a:} Mean value of the deviations of the binned oxygen abundance around the obtained broken O/H -- $R$ relation as a function of the position of the centre of the first bin. 
  {\em Panel} {\bf b:} Derived value of the central oxygen abundance  as a function of the position of the first bin centre. The bars show the uncertainties of the obtained values.   
  {\em Panel} {\bf c:} Same as panel (b), but for the abundances at the isophotal radius. 
  {\em Panel} {\bf d:} Value of the break radius  as a function of the position of the centre of the first bin. The circles denote the break radius in O/H -- $R$ distribution, and
                       the crosses mark the break radius in N/H -- $R$ distrinutions.  The bars show the uncertainties of the obtained values. 
}
\label{figure:rbin-oh-rb}
\end{figure}

\begin{figure}
\resizebox{1.00\hsize}{!}{\includegraphics[angle=000]{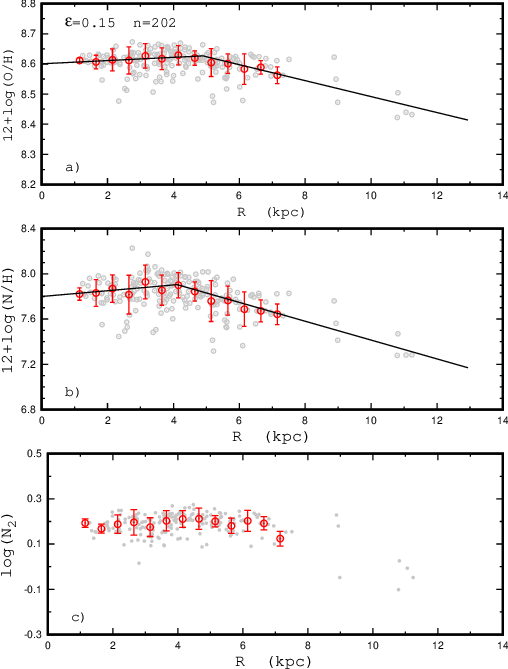}}
\caption{Abundance properties of NGC~3521 traced by the fibres where the line flux errors $\varepsilon <$ 0.15.
  {\em Panel} {\bf a:} Radial oxygen abundance distribution. The grey circles denote the $R$ calibration-based abundances for the individual fibres.
  The red circles denote the median values of the oxygen abundances in bins of 0.5 kpc in radius, and the bars show the scatter in O/H 
  around the median value of each bin. 
  The line shows the O/H -- $R$ relation for the case of  $\varepsilon <$ 0.30  (from panel (a) of Fig.~\ref{figure:r-oh-nh}). 
  {\em Panel} {\bf b:} Same as panel (a), but for the nitrogen abundances. 
  {\em Panel} {\bf c:} Intensity of the emission nitrogen line N$_{2}$ as a function of radius. The notations are as in panel (a). 
}
\label{figure:r-oh-nh-e-0.15}
\end{figure}

\begin{figure}
\resizebox{1.00\hsize}{!}{\includegraphics[angle=000]{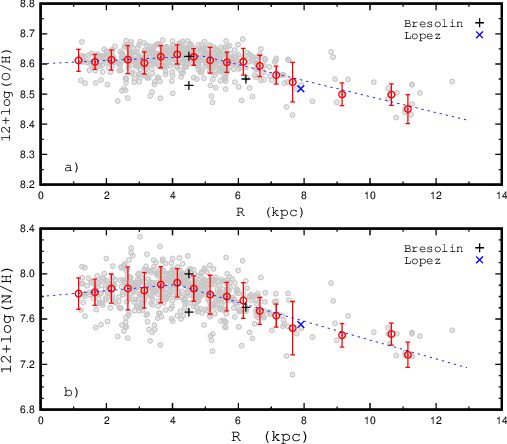}}
\caption{Comparison between abundances in NGC~3521 based on fibre measurements here and abundances based on  H\,{\sc ii} regions  from the literature. 
  {\em Panel} {\bf a:} Radial oxygen abundance distribution. The grey circles denote the $R$ calibration-based abundances for the individual fibres.
  The red circles denote the median values of the oxygen abundances in bins of 0.5 kpc in radius, and the bars show the scatter in O/H 
  around the median value of each bin.   The line shows the O/H -- $R$ relation for those data.
  The plus signs denote the abundances in H\,{\sc ii} regions measured by \citet{Bresolin1999}. The cross marks the abundance in H\,{\sc ii} region measured by \citet{Lopez2019}. 
  {\em Panel} {\bf b:} Same as panel (a), but for the nitrogen abundances. 
}
\label{figure:r-oh-nh-previous}
\end{figure}

\begin{figure}
\resizebox{1.00\hsize}{!}{\includegraphics[angle=000]{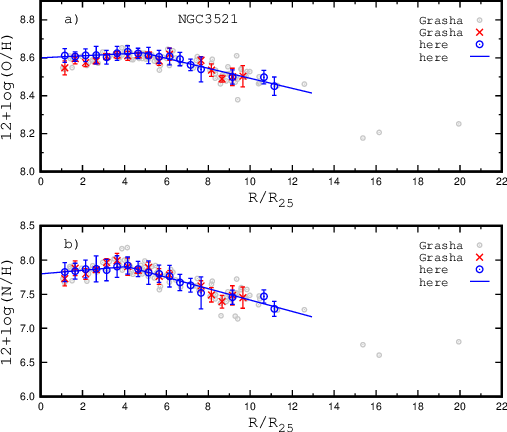}}
\caption{Comparison between abundances in NGC~3521 based on fibre measurements here and abundances based on  H\,{\sc ii} regions  from \citet{Grasha2022}. 
  {\em Panel} {\bf a:} Radial oxygen abundance distribution. The grey circles denote the $R$ calibration-based abundances for the individual  H\,{\sc ii} regions  from \citet{Grasha2022},  
  the red crosses denote the median values of the oxygen abundances in bins of 0.5 kpc in radius, and the bars show the scatter in O/H 
  around the median value of each bin.  The blue circles denote the median values of the oxygen abundances in bins for our fibre measurements, the line shows the O/H -- $R$
  relation for those data (the same as in Fig.~\ref{figure:r-oh-nh}). 
  {\em Panel} {\bf b:} Same as panel (a), but for the nitrogen abundances. 
}
\label{figure:r-oh-nh-grasha}
\end{figure}

The precision of the line flux measurements is specified by the ratio of the flux error to the flux, $\varepsilon$. We selected a sample of fibre spectra for which the parameter $\varepsilon$
is less than the adopted limit value for each line necessary for the abundance determination using the R calibration from \citet{Pilyugin2016}. The necessary emission lines include the oxygen
doublet [O\,{\sc ii}]$\lambda$3727,$\lambda$3729, the H$\beta$ line, the oxygen line [O\,{\sc iii}]$\lambda$5007, the H$\alpha$ line, and the nitrogen line [N\,{\sc ii}]$\lambda$6584.
The oxygen doublet [O\,{\sc ii}]$\lambda$3727,$\lambda$3729 is from the blue spectrum, while the remaining lines are from the red spectrum. To ensure consistency between the line fluxes in the
blue and red spectra, we also required that the absolute value of the relative discrepancy between the H$\beta$ fluxes in the red and blue spectra
(H$\beta_{blue}$ -- H$\beta_{red}$)/(0.5 (H$\beta_{blue}$ + H$\beta_{red}$)) is  smaller than the adopted  $\varepsilon$ value. 

Since the [O\,{\sc iii}]$\lambda$5007 and $\lambda$4959 lines originate from transitions at the same energy level,  their flux ratio is governed only by the transition probability ratio, 
which is close to 3 \citep{Storey2000}. Therefore, the value of R$_3$ can be well approximated as R${_3}$  = 1.33$\times I_{{\rm [OIII]} \lambda 5007}$/$I_{{\rm H}\beta }$. Given that the stronger
[O\,{\sc iii}]$\lambda$5007 line is usually measured with higher precision than the weaker [O\,{\sc iii}]$\lambda$4959 line, we estimated the value of R$_3$ in that way, and not as the sum
of the line fluxes R$_3$  = $I_{{\rm [O\,III]} \lambda 4959+ \lambda 5007} /I_{{\rm H}\beta }$. Similarly, the [N\,{\sc ii}]$\lambda$6584 and $\lambda$6548 lines also originate from transitions at the
same energy level, and the transition probability ratio for those lines is again close to 3 \citep{Storey2000}. Consequently, we also estimated the value of
N$_2$ as N$_{2}$  = 1.33$\times  I_{{\rm [NII]} \lambda 6584} /I_{{\rm H}\beta }$. 

We initially adopted a limit value of $\varepsilon$ = 0.3, which is satisfied by the required emission lines in the spectra of 926 fibres out of 2214 fibres. The measured line fluxes in the
spectra are corrected for   interstellar reddening using the reddening law of \citet{Cardelli1989} with $R_{V}$ = 3.1. The logarithmic extinction at H$\beta$ is estimated through a comparison
between the measured and theoretical $F_{{\rm}H\alpha}/F_{{\rm H}\beta}$ ratios where the theoretical value of the line ratio (= 2.87) is taken from \citet{Osterbrock2006}, assuming the case $B$
recombination. If the measured value of the  $F_{{\rm}H\alpha}/F_{{\rm H}\beta}$ ratio is lower than the theoretical value, then a reddening value of zero is adopted.

We classified the excitation of each spectrum using its position on the standard diagnostic Baldwin-Phillips-Terlevich (BPT) diagram [N\,{\sc ii}]$\lambda$6584/H$\alpha$ versus the
[O\,{\sc iii}]$\lambda$5007/H$\beta$, suggested by \citet{Baldwin1981}. As in our previous studies \citep{Zinchenko2019,Pilyugin2020,Pilyugin2021}, the spectra located to the left of (below)
the demarcation line of \citet{Kauffmann2003} are referred to as SF-like or H\,{\sc ii} region-like spectra (blue symbols in panel (a) of Fig.~\ref{figure:bpt}); those located to the right of
(above) the demarcation line of \citet{Kewley2001} are referred to as  active galactic nucleus (AGN)-like spectra (dark symbols in panel (a) of Fig.~\ref{figure:bpt}); and the spectra located
between the two demarcation lines are classified as intermediate (INT) spectra (red symbols in panel (a) of Fig.~\ref{figure:bpt}). Panel (b) of Fig.~\ref{figure:bpt} shows the distribution of
the individual fibres of different BPT types across the image of  NGC~3521.

The oxygen and nitrogen abundances can be determined for fibres with the H\,{\sc ii} region-like spectra using the R calibration from \citet{Pilyugin2016}. In our previous  paper
\citep{Pilyugin2022}, we tested the applicability of R calibration  to the IFU spectra. We estimated the calibration-based abundances obtained using both the IFU and the slit spectroscopy for
eight nearby galaxies. We find that the IFU and the slit spectra-based abundances obtained through the R calibration are close to each other; the mean value of the differences of binned
abundances is within 0.01 dex. We also found that the R calibration can produce close estimations of the abundances using IFU spectra obtained with different spatial resolutions and different
spatial samplings.
  
While the BPT diagram is a commonly used tool to identify the ionising source of nebular emission, the reliability of classifying the ionising source using only the BPT diagram has been
questioned \citep{CidFernandes2010,CidFernandes2011,Sanchez2014,Sanchez2021,Sanchez2024,Lacerda2018,D'Agostino2019,Johnston2023}. It has been suggested to use the equivalent width of the
H$\alpha$ emission line or the gas velocity dispersion as a diagnostic indicator in addition to the emission-line ratios that form the basis of the BPT diagnostic diagram. It has been
argued \citep[e.g.][]{Sanchez2024} that the ionisation in objects with a $EW_{{\rm H}\alpha} <$ 6 {\AA} is due to hot, old, low-mass, evolved stars (post-AGBs), which are associated with retired
regions in galaxies where no star formation occurs. Thus, fibres classified as H\,{\sc ii} region-like spectra based on the BPT diagram but with $EW_{{\rm H}\alpha} <$ 6 {\AA} are excluded from
the abundance determinations. It should be noted that only 5 out of the  578 fibres  classified as H\,{\sc ii} region-like have $EW_{{\rm H}\alpha} <$ 6 {\AA}. 

We estimated the oxygen abundances in 573 fibres using the $R$ calibration from \citet{Pilyugin2016}. We also estimated the nitrogen-to-oxygen abundance ratios using the
corresponding N/O calibration from \citet{Pilyugin2016} and then determined the nitrogen abundances using the relation log(N/H) = log(O/H) + log(N/O).
The radial distributions of  oxygen and nitrogen abundances for the individual spectra in the galaxy NGC~3521 are shown by grey points  in panels (a) and (b), respectively, of 
Fig.~\ref{figure:r-oh-nh}.  To minimise the effect of fibres with unreliable abundances in determining the radial abundance distribution, we used the median
values of the abundances in bins of 0.5 kpc in radius  (red circles in panels (a) and (b) of Fig.~\ref{figure:r-oh-nh}). Only bins with three or more points are considered. 
The binned O/H -- $R$ relation reveals a broken linear trend, described by the  equations
\begin{eqnarray}
       \begin{array}{llcl}
12+{\rm log(O/H)}  & = & 8.600      & +  0.0053 \times R,  \;\;\;  R < 4.9     \\
                   &   & (\pm0.007) &    (\pm0.0023),                            \\
                   & = & 8.756      & -  0.0264 \times R,  \;\;\;  R > 4.9     \\
                   &   & (\pm0.019) &    (\pm0.0024),                            \\
     \end{array}
\label{equation:oh}
\end{eqnarray}
shown by the line in panel (a) of Fig.~\ref{figure:r-oh-nh}.
The broken linear N/H -- $R$ relation 
\begin{eqnarray}
       \begin{array}{llcl}
12+{\rm log(N/H)}  & = & 7.799 & + 0.0255 \times R,  \;\;\;  R < 4.1     \\
                   &   & (\pm0.020) &    (\pm0.0079),                     \\
                   & = & 8.245 &- 0.0831 \times R,  \;\;\;   R > 4.1     \\
                   &   & (\pm0.052) &    (\pm0.0070),                     \\
     \end{array}
\label{equation:nh}
\end{eqnarray}
is shown by the line in panel (b) of Fig.~\ref{figure:r-oh-nh}. It should be noted that the abundance measurements are not available at the very centre of  NGC~3521 (within $\sim$1 kpc)
and near the optical radius (within $\sim$1.5 kpc) (see Fig.~\ref{figure:r-oh-nh}). Then the central abundance and abundance at the optical radius are extrapolated values, determined
using Eq.~\ref{equation:oh}  and Eq.~\ref{equation:nh}. 
 
The coefficients in  Eq.~\ref{equation:oh}  (and Eq.~\ref{equation:nh}) and the value of the break radius are derived using an iterative procedure.  In the first step, the value
of $R_{b}$ is chosen by visual inspection of the radial abundance distribution, and the coefficients are determined  by the best fit. Then a new value of the break radius is determined
as the radius where the abundance estimated using the relation for the inner part is equal to the abundance obtained with the relation for the outer part, and the new coefficients are
determined  by the best fit.  The iteration is stopped when the obtained values of  $R{_b}$   in successive steps agree within 0.01 kpc. The uncertainty of the break radius is estimated
in the following way. A random noise ($\alpha \times$ uncertainty of coefficient) is added to each coefficient in Eq.~\ref{equation:oh} (and Eq.~\ref{equation:nh}), where $\alpha$ is a
random number from --1 to 1. The value of the break radius is determined with the disturbed coefficients and its difference with the break radius for the undisturbed coefficients $dR_{b}$
is obtained. The mean difference $\left(\frac{1}{n}\sum_{j=1}^{n} (dR_{b,j})^{2}\right)^{1/2}$ for 1000 simulations is considered as the uncertainty in the break radius. We found that the
uncertainty in the break radius $R_{b,OH}$ in the O/H -- $R$ relation is 0.5 kpc and the uncertainty in the break radius $R_{b,NH}$ in the N/H -- $R$ relation is 0.4 kpc.  

The validity of the shape of the oxygen abundance gradient in the galaxy can be verified by comparing it with the variation in the nitrogen emission line N$_{2}$ intensity across the disc
of the galaxy \citep{Pilyugin2024}.  The standard notation for  nitrogen line intensities $N_2  = I_{\rm [N\,II] \lambda 6548+ \lambda 6584} /I_{{\rm H}\beta }$ is used. The intensity of the N$_{2}$
line correlates with the electron temperature in the nebula \citep[e.g.][]{Pilyugin2016} and, consequently, with oxygen abundance. In particular, it is assumed within the framework of the
N$_2$ calibration that the oxygen abundance is a function solely of the intensity of the N$_{2}$ line \citep{Pettini2004,Marino2013}. It should be emphasised that the value of N$_{2}$ alone
does not provide an accurate abundance determination of   H\,{\sc ii} regions, and the N/O ratio and the excitation parameter must also be considered \citep{Pilyugin2016, Schaefer2020}. 
Panel (c) in Fig.~\ref{figure:r-oh-nh} shows the variation in N$_{2}$ intensity across the disc of NGC~3521. The similarity between the radial oxygen abundance distribution (panel (a) in
Fig.~\ref{figure:r-oh-nh}) and the radial variation in N$_{2}$ intensity (panel (c) in Fig.~\ref{figure:r-oh-nh}) confirms the validity of the shape of the oxygen abundance gradient in NGC~3521.

To examine the influence of the chosen positions of the centres of the bins on the derived parameters of the oxygen and  nitrogen abundance distributions, we considered these gradients
for ten positions of the centres of the bins. The position of the centre of $j$-th bin, $R_{cb,j}$,  is $R_{cb,j}$ = $R_{cb,1}$ + $(j-1)\times$(bin size), where the bin size is equal to 0.5 kpc.
We considered ten variants of the position of the first bin centre from  $R_{cb,1}$ = 0 to  $R_{cb,1}$ = 0.45 kpc with the step of 0.05 kpc. Figure~\ref{figure:rbin-oh-rb} shows the derived
parameters of the abundances distributions as a function of the position of the first bin centre: the mean value of the deviations of the binned oxygen abundances around the obtained broken
O/H -- $R$ relation (panel (a)), the derived values of the central oxygen abundance (panel (b)), the oxygen abundance at the optical radius (panel (c)), and the values of the break radii
in the O/H -- $R$ distributions (circles) and in the N/H -- $R$ distributions (crosses) (panel (d)). Panel (b) in Fig.~\ref{figure:rbin-oh-rb} shows that the derived value of the central
oxygen abundance depends weakly on the adopted positions of the centres of the bins; the difference between the values of (O/H)$_{0}$ for  any two values of the $R_{cb,1}$ is within 0.025 dex.   
Similarly, panel (c) shows that the derived value of the oxygen abundance at the optical radius also depends weakly on the adopted positions of the centres of the bins; again, the difference
between the values of (O/H)$_{R_{25}}$ for any two values of $R_{cb,1}$ is within 0.025 dex. Panel (d) of Fig.~\ref{figure:rbin-oh-rb}  shows that the values of the break radius in the O/H -- $R$
distributions estimated for different $R_{cb,1}$ can differ by more than 1 kpc, from $R_{b,OH}$ = 4.56 kpc for $R_{cb,1}$ = 0.35 kpc to $R_{b,OH}$ = 5.77 kpc for $R_{cb,1}$ = 0.05 kpc. The values
of the break radius in the N/H -- $R$ distributions estimated for different $R_{cb,1}$ can change within 0.5 kpc, from $R_{b,NH}$ = 4.05 kpc for $R_{cb,1}$ = 0.25 kpc to $R_{b,OH}$ = 4.46 kpc for
$R_{cb,1}$ = 0.05 kpc. It should be noted that the $R_{b,NH}$ is smaller than the $R_{b,OH}$ for each value of the  $R_{cb,1}$, but the difference is less than the uncertainties of these values.  
Panel (a) in Fig.~\ref{figure:rbin-oh-rb} shows that the mean value of the deviations of the binned oxygen abundances around the obtained broken O/H -- $R$ relation is between $\sigma_{OH}$ =
0.010 and  $\sigma_{OH}$ = 0.018 dex, with one exception for the case of  $R_{cb,1}$ = 0.20 kpc. Thus, the abundances at the centre and optical radius depend weakly on the bin position,  while
the breaking radii and their uncertainties can exhibit greater variability. Notably, $R_{b,OH}$ $>$ $R_{b,NH}$ for any tested bin position, although the differences are comparable to the
uncertainties. We use the distribution with minimum value of the $\sigma_{OH}$ (the case of $R_{cb,1}$ = 0.15 kpc,  $\sigma_{OH}$ = 0.0105 dex) in our discussion of the abundances distribution in
the NGC 3521. These distributions are shown in Fig.~\ref{figure:r-oh-nh} and are approximated by Eq.~\ref{equation:oh}  and Eq.~\ref{equation:nh}. 

To assess the effect of the chosen $\varepsilon$ limit on the derived oxygen and nitrogen abundance distributions, we computed these gradients using a stricter limit of $\varepsilon$ = 0.15. 
The  necessary emission lines in 202  H\,{\sc ii} region-like spectra satisfy the condition $\varepsilon$ $\la$ 0.15. The grey points in panel (a) of Fig.~\ref{figure:r-oh-nh-e-0.15} denote
the oxygen abundances in individual fibres, and the red circles indicate the binned values of the abundances for fibres selected using  $\varepsilon \la$ 0.15. The line in panel (a)
of Fig.~\ref{figure:r-oh-nh-e-0.15} shows the O/H -- $R$ relation for $\varepsilon <$ 0.30  (from panel (a) of Fig.~\ref{figure:r-oh-nh}, Eq.~\ref{equation:oh}). Panel (b) of
Fig.~\ref{figure:r-oh-nh-e-0.15} shows a similar diagram for the nitrogen abundances. Both panels show that binned values for $\varepsilon$ $\la$ 0.15 closely follow the O/H -- $R$ and
N/H -- $R$ relations derived with $\varepsilon$ $\la$ 0.30. This consistency suggests that measurement errors in the fibre spectra are random. Thus, a sample of fibres selected with
$\varepsilon \la$ 0.3 provides a reliable estimate of the radial oxygen and nitrogen abundance distributions in the disc of NGC~3521.

The spectra of the H\,{\sc ii} regions in NGC~3521 were measured by \citet{Zaritsky1994}, \citet{Bresolin1999}, and \citet{Lopez2019}. Unfortunately,  \citet{Zaritsky1994} did not measure
the fluxes of the nitrogen emission line [N\,{\sc ii}]$\lambda$6584 required for the abundance determination through the R calibration. We estimated the oxygen and nitrogen abundances for
three H\,{\sc ii} regions using the line measurements from \citet{Bresolin1999}. The galactocentric distances of those H\,{\sc ii} regions were determined using the $R/R_{25}$ values reported
in \citet{Zaritsky1994}. These H\,{\sc ii} regions are overlaid as plus signs on the O/H -- R (panel (a) of Fig.~\ref{figure:r-oh-nh-previous}) and on the N/H -- R (panel (b) of
Fig.~\ref{figure:r-oh-nh-previous}) diagrams. \citet{Lopez2019} investigated an H\,{\sc ii} region near an ultraluminous X-ray source (ULX) in  NGC~3521, located $\sim$105 pc to the NE;
this distance was rescaled adopting the distance to NGC 3521 used  here, d = 10.7 Mpc, from the d=14.29 Mpc used by \citet{Lopez2019}. We  estimated the oxygen and nitrogen abundance
in this  H\,{\sc ii} region through the R calibration using their line measurements. The galactocentric distance of the H\,{\sc ii} region is determined using the ULX source coordinates
reported by \citet{Lopez2017}. This H\,{\sc ii} region is overlaid as a cross on the O/H -- R (panel (a) of Fig.~\ref{figure:r-oh-nh-previous}) and on the N/H -- R (panel (b) of
Fig.~\ref{figure:r-oh-nh-previous}) diagrams. Figure~\ref{figure:r-oh-nh-previous} shows that the abundance in H\,{\sc ii} regions measured by \citet{Bresolin1999} and by \citet{Lopez2019}
align with the distribution of abundances derived   from the fibre spectra in this study. 

It was noted above that the spectra of 92 H\,{\sc ii} regions in NGC~3521 were measured by \citet{Grasha2022}. So the fluxes of the emission lines required for the abundance determination
through the R calibration are available. We estimated the oxygen and nitrogen abundances for those H\,{\sc ii} regions using the line measurements from \citet{Grasha2022}. The galactocentric
distances of those  H\,{\sc ii} regions were determined using the position angle of the major axis and the inclination angle obtained here. The R calibration-based oxygen abundance as a
function of the radius for individual H\,{\sc ii} regions is shown by the grey circles in panel (a) of Fig.~\ref{figure:r-oh-nh-grasha}. The red crosses denote the median values of the oxygen
abundances in bins of 0.5~kpc in radius, and the bars show the scatter in O/H  around the median value of each bin. The blue circles denote the median values of the oxygen abundances in bins
for our fibre measurements; the line shows the O/H -- $R$  relation for those data (from Fig.~\ref{figure:r-oh-nh}). A close inspection of panel (a) of Fig.~\ref{figure:r-oh-nh-grasha} shows
that the binned oxygen abundances for the H\,{\sc ii} regions measured by \citet{Grasha2022} and the binned oxygen abundances for the fibres measured here agree within the uncertainties. 
Panel (b) of Fig.~\ref{figure:r-oh-nh-grasha} shows a similar comparison for nitrogen abundances. A close examination of panel (b) of Fig.~\ref{figure:r-oh-nh-grasha} shows again that the
binned nitrogen abundances for the H\,{\sc ii} regions measured by \citet{Grasha2022} and the binned nitrogen abundances for the fibres measured here agree within the uncertainties. 

Thus, the oxygen abundance in the inner region of the disc of  NGC~3521 (within $R_{b,OH}$) remains nearly constant, while the oxygen abundance gradient is negative at larger radii (panel (a)
of Fig.~\ref{figure:r-oh-nh} and Eq.~\ref{equation:oh}). The nitrogen abundance shows a similar pattern. The break in the O/H distribution occurs at a smaller radius than the O/H distribution
break (panel (b) of Fig.~\ref{figure:r-oh-nh} and Eq.~\ref{equation:nh}), but the difference between $R_{b,OH}$ and  $R_{b,NH}$ is within the uncertainties of these values. Some spiral galaxies
show a similar behaviour of the oxygen and nitrogen abundances with radius \citep[e.g.][]{Kreckel2019,Pilyugin2024}. \citet{Kreckel2019}  have determined the gas phase oxygen abundances for
H\,{\sc ii} regions across the discs of eight nearby galaxies using Very Large Telescope/Multi Unit Spectroscopic Explorer (MUSE) optical integral field spectroscopy as part of the PHANGS-MUSE
survey. One of these galaxies (NGC~4254)  shows a flattening in the oxygen abundance distribution in the inner part of the disc with the break radius of $R_{b,OH}$ $\sim$0.2$R_{25}$.  
A sample of 60 well-measured spiral galaxies from the MaNGA survey with different shapes of  radial oxygen abundance distributions was examined in \citet{Pilyugin2024}. It was found that 23
of the 60 galaxies belong to the sequence of galaxies with  level-slope (LS) gradients, where the oxygen abundance in the inner region of the disc is nearly constant  (the gradient is flatter
than $-0.05$~dex/$R_{25}$) and decreases at larger radii. The scatter of the binned oxygen abundances around the broken O/H--$R$ relation is smaller than $\sim 0.01$~dex. The radial nitrogen
abundance distributions in the majority of galaxies with LS-type distributions  show breaks at  smaller radii than the O/H distribution breaks.

It was shown in \citet{Pilyugin2024} that such observed behaviour of oxygen and nitrogen abundances, constant levels in the inner disc and negative gradients at larger radii, with nitrogen
breaks occurring at smaller radii than oxygen, can be naturally explained by accounting for the time delay between nitrogen and oxygen enrichment \citep[see e.g. Fig. 1 in][]{Maiolino2019},
and the variation in the star formation history along the radius predicted by the inside-out disc evolution model \citep[e.g.][]{Matteucci1989}. The nearly constant oxygen abundance in the
inner region $R < R_{b,{\rm OH}}$, panel (a) in Fig.~\ref{figure:r-oh-nh}, suggests that after the region has reached a high astration level some time ago $\tau(R)$, the star formation rate in
the region decreases. The short-living massive stars in this region will produce relatively small amounts of oxygen. At the same time the production of nitrogen by long-living low-mass stars
from the previous generations of stars continues. The increase in the nitrogen abundance as the radius decreases in the zone of constant oxygen abundance (within  $R_{b,{\rm OH}}$ $\ga$ $R$ $\ga$
$R_{b,{\rm NH}}$) suggests that the number of nitrogen-producing stars from previous generations, which have enough time to complete their evolution and eject their nitrogen into the interstellar
medium, increases as the radius decreases; that is, the value of  $\tau(R)$ increases as the radius decreases. The value of $\tau(R)$ becomes high enough at  radius $R_{b,{\rm NH}}$  that a bulk
of  nitrogen-producing stars from previous generations complete their evolution;  the nitrogen abundance then reaches a high level at $R_{b,{\rm NH}}$ and  remains nearly constant at
$R < R_{b,{\rm NH}}$. Thus, the observed behaviour of oxygen and nitrogen abundances with radius in  NGC~3521 clearly supports the  inside-out disc evolution model, where the galactic centre of
the galaxy evolves more rapidly than the outer regions  \citep{Matteucci1989}. The radial distributions of  oxygen and nitrogen abundances  obtained here provide evidence that  NGC~3521 is a
typical LS-gradient galaxy.

\section{Comparison between NGC~3521 and the Milky Way}

A comparison between the global structural characteristics of  NGC~3521 and the Milky Way is given in \citet{McGaugh2016} and \citet{Pilyugin2023}. These studies reveal a remarkable agreement
between the key parameters in the two galaxies: the optical radius of the MW is $R_{25}$ = 12 kpc  versus $R_{25}$ = 12.94 kpc for  NGC~3521; the stellar mass of the MW is log$M_{\star}/M_{\sun}$ =
10.716  versus log($M_{\star}/M_{\sun}$) = 10.70  for  NGC~3521; the rotation velocity of the MW is $V_{rot}$ = 235 km\,s$^{-1}$  versus  $V_{rot}$ = 227~km\,s$^{-1}$ for  NGC~3521; the mass of the
central supermassive black hole in the MW is log($M_{BH}/M_{\sun}$) = 6.618  versus log($M_{BH}/M_{\sun}$) = 6.85$\pm$0.58  in  NGC~3521. This illustrates that the galaxy NGC~3521 has  structural
parameters that are very similar to those of the MW, which means that   NGC~3521 is a nearly structural analogue of the Milky Way. Here we compare the radial distributions of the oxygen abundances,
the gas mass fractions, and the effective oxygen yields (indicators of the evolutionary stage) of  NGC~3521 and the Milky Way.

\subsection{Oxygen abundance}

\begin{figure}
\resizebox{1.00\hsize}{!}{\includegraphics[angle=000]{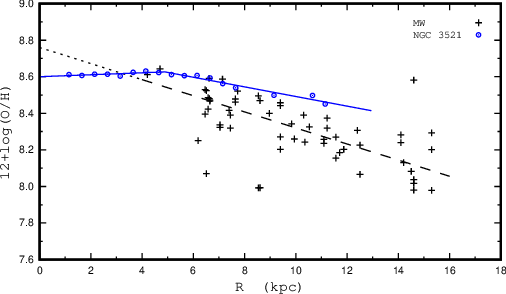}}
\caption{Comparison between radial oxygen abundance distributions in  NGC~3521 and the Milky Way.
  The plus signs denote the abundances in the  H\,{\sc ii} regions of the Milky Way, and the dashed line is the O/H--$R$ relation adopted for the Milky Way.
  The circles mark the binned abundances in  NGC~3521 (from panel (a) of Fig.~\ref{figure:r-oh-nh}). 
  The solid line shows the  O/H--$R$ relation adopted for  NGC~3521,  Eq.~\ref{equation:oh}. 
}
\label{figure:c-r-oh}
\end{figure}

\begin{figure*}
\resizebox{1.00\hsize}{!}{\includegraphics[angle=000]{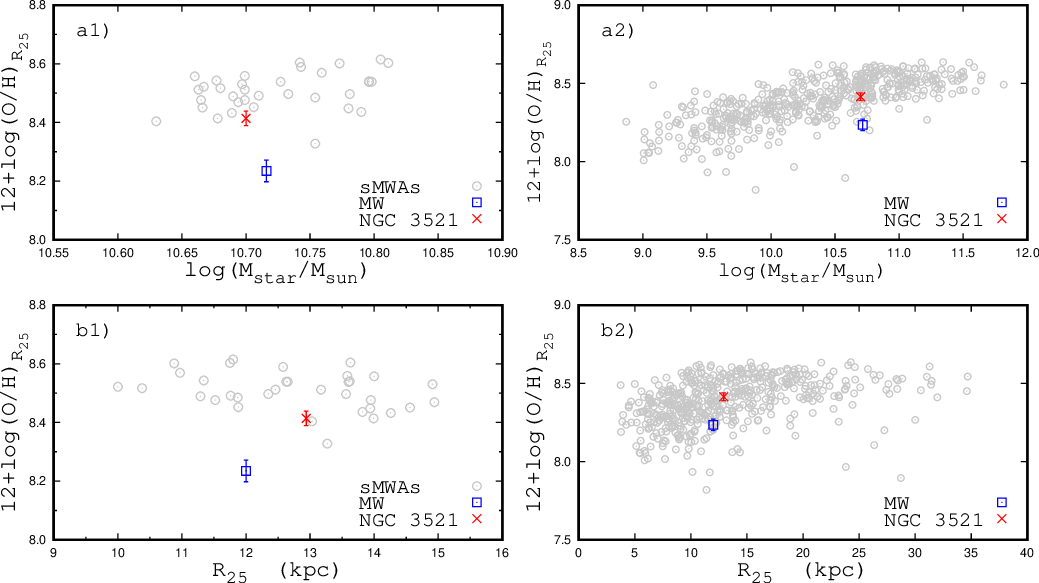}}
\caption{Comparison of the oxygen abundance at the optical radius in the Milky Way and in the structural Milky Way analogues.
    {\sl Panel {\bf a1}:} Oxygen abundance at the optical radius of the galaxy as a function of its stellar mass. The grey circles designate the structural Milky Way analogues,
                         the square marks the Milky Way, and the cross is the NGC~3521. 
    {\sl Panel {\bf a2}:} Same as panel (a1), but the grey circles show a total sample of galaxies from \citet{Pilyugin2023} (not only the structural Milky Way analogues). 
    {\sl Panel {\bf b1}:}  Oxygen abundance at the optical radius of the galaxy as a function of the optical radius. The designations are the same as in panel (a1). 
    {\sl Panel {\bf b2}:} Same as panel (b1), but the grey circles show a total sample of galaxies. 
  }
\label{figure:c-m-r25-oh25}
\end{figure*}

The radial distribution of the gas-phase oxygen abundances in the Milky Way  was determined in a number of investigations \citep[e.g.][]{Shaver1983,Deharveng2000,Rudolph2006,Fernandez2017,
Arellano2020,Arellano2021}. The electron temperatures and ionic abundances in the MW H\,{\sc ii} regions are usually determined using the measurements of the optical emission lines. 
However, the observations of the radio recombination lines were used in the determination of the electron temperatures in the  H\,{\sc ii} regions by \citet{Shaver1983} and \citet{Deharveng2000},
and the measurements of the far-IR emission lines were used in the ionic abundances determinations by \citet{Rudolph2006}. A mandatory condition of the current study is that the abundances
in the MW and in  NGC~3521 are determined using the same metallicity scale. We considered a sample of 41 H\,{\sc ii} regions of the MW from \citet{Arellano2020} and \citet{Arellano2021}.
The electron temperatures and ionic abundances in these H\,{\sc ii} regions were determined using the measurements of the optical emission lines. However, the original  $T_{e}$-based oxygen
abundances of the H\,{\sc ii} regions were not used in this study; instead, they were re-estimated for the following reason: the oxygen abundances in the H\,{\sc ii} regions of the MW are
being compared to those of NGC~3521 determined through the $R$ calibration from \citet{Pilyugin2016}. The $T_{e}$-based oxygen abundances of the H\,{\sc ii} regions, used as  calibrating data
points in the construction of the $R$ calibration, were derived using the $T_{e}$-method equations reported in \citet{Pilyugin2012}. To ensure that the Milky Way oxygen abundances correspond
to the same metallicity scale as the $R$ calibration-based abundances in NGC 3521, the oxygen abundances in the Milky Way H\,{\sc ii} regions were re-calculated using the same $T_{e}$-method
equations from \citet{Pilyugin2012}.

The line intensity measurements in the spectra of H\,{\sc ii} regions are from the same sources as those in \citet{Arellano2020} and \citet{Arellano2021}. If the measurements of two
auroral lines ([O\,{\sc iii}]$\lambda$4363 and [N\,{\sc ii}]$\lambda$5755) are available for the H\,{\sc ii} region, then two values of the oxygen abundance are determined. 
The first value of the abundance is based on the electron temperature in the high-ionisation part of the nebula $t_{3}$ estimated using the auroral line [O\,{\sc iii}]$\lambda$4363,
while the electron temperature in the low-ionisation zone ($t_{2}$) is obtained through the canonical relationship between $t_{2}$ and $t_{3}$:  $t_{2}$ = 0.7$t_{3}$ + 0.3
\citep{Campbell1986,Garnett1992}. The second value of the abundance is based on the electron temperature in low-ionisation zone estimated from the measured auroral line
[N\,{\sc ii}]$\lambda$5755, and the electron temperature in high-ionisation zone is determined from the relationship between electron temperatures.

The oxygen abundances of individual H\,{\sc ii} regions in the Milky Way  are shown by the plus signs in  Fig.~\ref{figure:c-r-oh}. The adopted O/H -- $R$ relation in the Milky Way is given by
\begin{equation}
12+\log(\rm O/H) = 8.762(\pm0.043) -0.0439(\pm0.0042) \times R  
\label{equation:mw-oh-r}
\end{equation}
where $R$ is the radius in kiloparsecs. This relation is shown by the dashed line  in Fig.~\ref{figure:c-r-oh} in the radius range from 4 kpc to 16 kpc (where the measurements of the spectra of
H\,{\sc ii} regions are available), an extrapolation of this relation to the centre of the Milky Way is shown by the dotted line. This O/H -- $R$ relation gives an oxygen abundance at the optical
radius ($R_{25}$ = 12 kpc) of 12 + log(O/H)$_{R_{25}}$ = 8.235. The binned oxygen abundances in the disc of  NGC~3521 (from panel (a) of Fig.~\ref{figure:r-oh-nh}) are shown by circles in
Fig.~\ref{figure:c-r-oh}. The  O/H -- $R$ relation in  NGC~3521 (Eq.~\ref{equation:oh}) is shown by the solid line  in Fig.~\ref{figure:c-r-oh}, yielding an oxygen abundance of
12 + log(O/H)$_{R_{25}}$ = 8.414 at the optical radius ($R_{25}$ = 12.94 kpc) of  NGC~3521. 

Figure~\ref{figure:c-r-oh} shows that the oxygen abundances of the two H\,{\sc ii} regions nearest to the centre of the Milky Way (at galactocentric distances of 4-5 kpc) are close to the binned
oxygen abundances in  NGC~3521 at the same galactocentric distances. The lack of  abundance measurements in the central region of the Milky Way prevents an accurate determination of the oxygen
abundance at the centre. An extrapolation of the linear  O/H -- $R$ relation to the very centre of the Milky Way may be unjustified since the radial oxygen abundance distribution in the Milky Way
can show a flattening in the central part \citep[e.g.][]{Arellano2020}.  However, a definitive conclusion on the flattening of the oxygen abundance gradient in the central part of the Milky Way
can only be made  based on the  abundance determinations of  H\,{\sc ii} regions close to the centre. Thus, the oxygen abundance in  NGC~3521 is close to that of the Milky Way at the radii of
4-5 kpc (near the radius break in the O/H -- $R$ relation for  NGC~3521) and the oxygen abundance in  NGC~3521 is higher than in the Milky Way at larger radii.  

Panels (a1) and (b1) in Fig.~\ref{figure:c-m-r25-oh25} show the comparison between the oxygen abundance at the optical radius in the Milky Way and  the structural Milky Way analogues
from \citet{Pilyugin2023}. Panels (a1) and (b1) in Fig.~\ref{figure:c-m-r25-oh25} show that the oxygen abundances at the optical radius of the Milky Way are lower than those
in NGC~3521 and  any other structural Milky Way analogues. It should be noted that the oxygen abundances at the optical radii of some spiral galaxies (which are not  sMWAs) are lower
than in the Milky Way (panels (a2) and (b2) in Fig.~\ref{figure:c-m-r25-oh25}). For example, the R calibration-based oxygen abundance  at the optical radius in the well-studied giant
spiral galaxy M~101 \citep[e.g.][]{Kennicutt1996,Pilyugin2001,Kennicutt2003,Croxall2016} is 12 + log(O/H) $\sim$ 7.9, which is lower by a factor of around two than in the Milky Way. 
The oxygen abundance at the optical radius of  NGC~3521 is lower than that of most  sMWAs (panels (a1) and (b1) in Fig.~\ref{figure:c-m-r25-oh25}). The difference between the abundances
at the optical radius between  NGC~3521 and the Milky Way is smaller than that of most sMWAs.   NGC~3521, being the nearest structural analogue of the Milky Way, is also among the galaxies
that show the smaller differences to the Milky Way in the oxygen abundances at the optical radius, although the difference is as large as 0.17 dex.

Thus, to make a solid conclusion on the discrepancy (or similarity) between the central oxygen abundances in  NGC~3521 and the Milky Way, the abundance determinations of H\,{\sc ii} regions
near the centre of the Milky Way should be performed. The oxygen abundance in the Milky Way is closer to that of NGC~3521 at the radii of 4-5 kpc (near the radius break in the
O/H -- $R$ relation for  NGC~3521). The oxygen abundances in the outer part of the Milky Way are lower than those in the outer part of  NGC~3521; the difference in the oxygen abundances
at the optical radius is as large as $\sim$0.17 dex. The lower oxygen abundances in the outer part of the Milky Way in comparison to  NGC~3521 can be caused by the following: a higher gas mass
fraction in the outer part of the Milky Way, an inflow of a larger amount of low-metallicity gas,  and a higher efficiency of the heavy element loss through  Galactic winds in the outer part
of the Milky Way compared to the outer part of  NGC~3521. We compare  the radial distributions of the gas mass fraction  in the Milky Way and in  NGC~3521 in the following section.

\begin{table}
\caption[]{\label{table:gas}
  Parameters of the radial distributions  of the surface mass density of atomic H\,{\sc i} and molecular H$_{2}$ hydrogen in the Milky Way
  (the coefficients in  Eq.~\ref{equation:r-h}). 
}
\begin{center}
\begin{tabular}{lccc} \hline \hline
                     &
R$_{d}$               &
R$_{m}$               &
$\Sigma_{0}$          \\
                      &
(kpc)                 &
(kpc)                 &
(M$_{\sun}$/pc$^{2}$)   \\   \hline         
H\,{\sc i}+He   &   7   &   4   &  53.1      \\ 
H$_{2}$+He       &   1.5 &  12   &  2180      \\ 
                    \hline
\end{tabular}\\
\end{center}
\end{table}

\subsection{Gas mass fraction}

\begin{figure}
\resizebox{1.00\hsize}{!}{\includegraphics[angle=000]{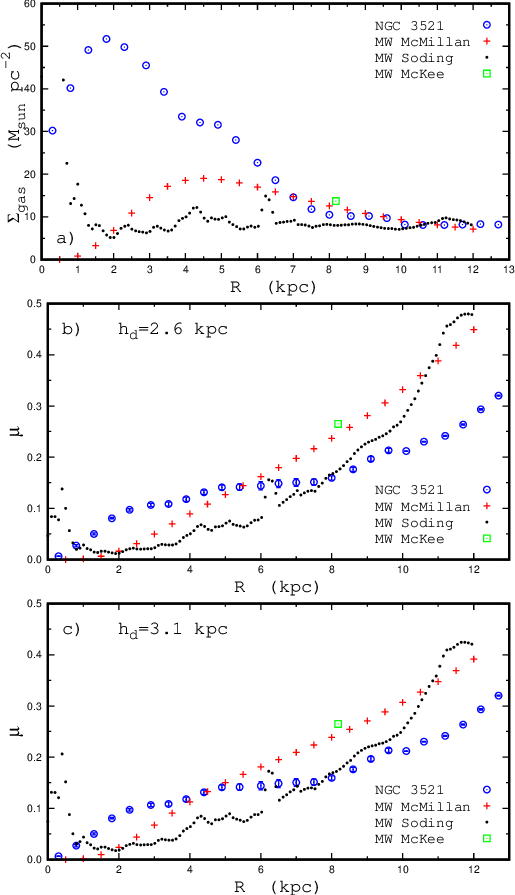}}
\caption{Comparison between the radial distributions of the gas mass surface density and the gas mass fraction in  NGC~3521 and the Milky Way.
  {\sl Panel {\bf a}:} Radial distributions of the gas mass surface density in  NGC~3521 and the Milky Way.
  The circles designate the gas mass surface densities in  NGC~3521.
  The plus signs show the gas mass surface densities in the Milky Way from \citet{McMillan2017}. 
  The points are the gas mass surface densities in the Milky Way from \citet{Soding2024}. 
  The square marks the gas mass surface density in the solar vicinity from \citet{McKee2015}. 
  {\sl Panel {\bf b}:} Radial distributions of the gas mass fraction in  NGC~3521 and the Milky Way. The  gas mass fractions in NGC~3521 are shown by the circles;
  the bars display the uncertainties (the bars are usually smaller than the symbol size).   The  gas mass fractions in the Milky Way were
  estimated with the stellar disc scale length of $h_{d}$ = 2.6 kpc and the gas mass surface densities shown in panel (a) . The notations are the same as in panel (a).  
  {\sl Panel {\bf c}:} Same as panel (b), but for the stellar disc scale length of $h_{d}$ = 3.1 kpc. 
}
\label{figure:c-r-mu}
\end{figure}

Here we compare the radial distributions of the  gas mass fraction $\mu$ or the astration level $s$ across the Milky Way and  NGC~3521. The value of the astration level $s$ indicates the
fraction of the gas converted into stars, $s$ = $M_{\star}/(M_{\star} + M_{gas})$,  that is $s$ = 1 - $\mu$. The mass surface densities of the atomic gas, molecular gas, and stars as a function
of the radius in  NGC~3521 were taken from \citet{Leroy2008}, where the surface densities are corrected for inclination and a factor of 1.36 is used in the determination of  atomic and
molecular gas densities to account for helium. The gas mass surface density as a function of  radius in  NGC~3521 is shown by the circles in panel (a) of Fig.~\ref{figure:c-r-mu}, and 
the gas mass fraction as a function of  radius is shown by the circles in panels (b) and (c) of Fig.~\ref{figure:c-r-mu}. The uncertainty of the obtained values of the gas mass fraction at
a given radius is estimated in the following way. A random noise ($\alpha \times$ uncertainty) is added to the mass surface densities of the atomic gas, molecular gas, and stars (whose values
and their uncertainties are from \citealt{Leroy2008}), where $\alpha$ is a random number from --1 to 1. The value of  $\mu$ is estimated from the disturbed surface densities, and its
difference with the gas mass fraction for the undisturbed surface densities is determined. The mean difference for 1000 simulations is considered as the uncertainty of the value of the gas
mass fraction. The obtained uncertainties of the values of the gas mass fraction in NGC~3521 are shown by bars in panels (b) and (c) of Fig.~\ref{figure:c-r-mu}. 

The radial distributions of the surface mass density of atomic  H\,{\sc i} and molecular H$_{2}$ hydrogen in the Milky Way are discussed in many investigations 
\citep[e.g.][]{Dame1993,Kalberla2008,Nakanishi2016,Marasco2017,McMillan2017,Bacchini2019,MertschPhan2023,Soding2024}. We considered two of the most different distributions for the surface gas
mass density in the Milky Way. The widely used radial distribution of the surface gas mass density from \citet{McMillan2017} is based on the investigations of \citet{Dame1993} and
\citet{Kalberla2008}, the distributions of the surface mass density of atomic H\,{\sc i} and molecular H$_{2}$ hydrogen (including helium) are approximated by the expression 
\begin{equation}
\Sigma_{H+He} = \Sigma_{0} \exp \left (-\frac{R_{m}}{R} - \frac{R}{R_{d}} \right )  .
\label{equation:r-h}
\end{equation}
The coefficients $\Sigma_{0}$, R$_{m}$, and R$_{d}$ for atomic H\,{\sc i} and molecular H$_{2}$ hydrogen are given in Table~\ref{table:gas}. The radial distribution of the gas mass surface
density in the Milky Way from \citet{McMillan2017} is shown by the plus signs in panel (a) of Fig.~\ref{figure:c-r-mu}. We also considered the radial distribution of the hydrogen surface
mass density from \citet{Soding2024}. We added the helium contribution (a factor of 1.36) to the original surface hydrogen mass density. The radial distribution of the gas mass surface
density in the Milky Way from \citet{Soding2024} is shown by the dark points in panel (a) of Fig.~\ref{figure:c-r-mu}. The gas mass surface density in the solar vicinity
(13.7$\pm$1.6 $M_{\sun}$ pc$^{-2}$) from \citet{McKee2015} is shown by the square in panel (a) of Fig.~\ref{figure:c-r-mu}. It should be noted that if the radial distribution obtained by
\citet{McMillan2017} is an adequate description of the gas mass surface density in the Milky Way, then the gas mass surface density in the solar vicinity is close to the azimuthally
averaged surface mass density of the gas (12.6 $M_{\sun}$ pc$^{-2}$) at the solar galactocentric distance. If the radial distribution obtained by \citet{Soding2024} adequately describes
the gas mass surface density in the Milky Way, then the gas mass surface density in the solar vicinity significantly exceeds the azimuthally averaged surface mass density of gas
(8.1 $M_{\sun}$ pc$^{-2}$) at the solar galactocentric distance.
  
The radial distribution of the surface stellar mass density in the Milky Way disc is described by the expression   
\begin{equation}
\Sigma_{\star} = \Sigma_{\star,R_{0}} \exp \left (-\frac{R - R_{0}}{h_{d}} \right )  ,
\label{equation:r-star}
\end{equation}
where  $\Sigma_{\star,R_{0}}$ is the surface stellar mass density at the solar galactocentric distance \citep[$R_{0}$ = 8.178$\pm$0.013 kpc, ][]{Gravity2019} and $h_{d}$ is the disc scale length.
\citet{Flynn2006} found a local stellar disc surface density of 35.5 $M_{\sun}$/pc$^{2}$, while \citet{McKee2015} found the value of 33.4 $M_{\sun}$/pc$^{2}$. The solar neighbourhood is located
in the interarm region, and  the `counted' local stellar disc surface density should be corrected for the spiral arm enhancement in order to find the azimuthal average of the surface stellar
mass density at the solar galactocentric distance, which means that  a 10\% enhancement should be added to the counted local stellar disc surface density \citep{Flynn2006,Kubryk2015}. This
results in a stellar disc surface density at the solar galactocentric distance of 39~$M_{\sun}$/pc$^{2}$ \citep{Flynn2006}  and 37~$M_{\sun}$/pc$^{2}$ \citep{McKee2015}. We adopted
$\Sigma_{\star,R_{0}}$ = 38~$M_{\sun}$/pc$^{2}$. \citet{BlandHawthorn2016} analysed 130 papers on disc parameters, with scale lengths ranging from 1.8 to 6.0~kpc. Their analysis of the main papers
(15 in all) on this topic led to $h_{d}$ = 2.6 $\pm$ 0.5~kpc.

The radial distribution of the gas mass fraction in the Milky Way obtained using the distribution of the surface stellar mass densities for the adopted disc scale length of 2.6 kpc and
the radial distribution of the surface gas mass density from \citet{McMillan2017},  is shown by the plus signs in  panel (b) of Fig.~\ref{figure:c-r-mu}. The radial distribution
of the gas mass fraction obtained using the radial distribution of the surface gas mass density from \citet{Soding2024}  is shown by the points. The square  in  panel (b) of
Fig.~\ref{figure:c-r-mu} marks the gas mass fraction in the solar vicinity estimated using the values of the gas mass surface density from \citet{McKee2015}. To examine the influence of
the disc scale length on the radial distribution of surface stellar mass densities and its effect on the radial distribution of gas mass fractions, we computed the radial distributions
of gas mass fraction for a disc scale length of  $h_{d}$ = 3.1 (panel (c) of Fig.~\ref{figure:c-r-mu}).

Figure~\ref{figure:c-r-mu} shows that the gas mass fraction in the Milky Way determined with the surface gas mass density from \citet{McMillan2017} is lower than the gas mass fraction of
NGC~3521 in the inner part (at radii $\la$5 kpc) and higher in the outer part  (at radii $\ga$5 kpc). The gas mass fraction in the Milky Way determined with the surface gas mass density
from \citet{Soding2024} is lower than the gas mass fraction of NGC~3521 within the galactocentric distances from $\sim$1 kpc to $\sim$8 kpc, and higher in the central part (at radii $\la$1
kpc) and in the outer part (at radii $\ga$8 kpc). In the next section we examine whether the discrepancy in oxygen abundances between the outer parts of the Milky Way and NGC~3521 can be
attributed to a difference in the gas mass fraction only or if variations in the efficiency of gas exchange with the surroundings also contribute to the O/H discrepancy.

\subsection{Effective oxygen yield}

\begin{figure}
\resizebox{1.00\hsize}{!}{\includegraphics[angle=000]{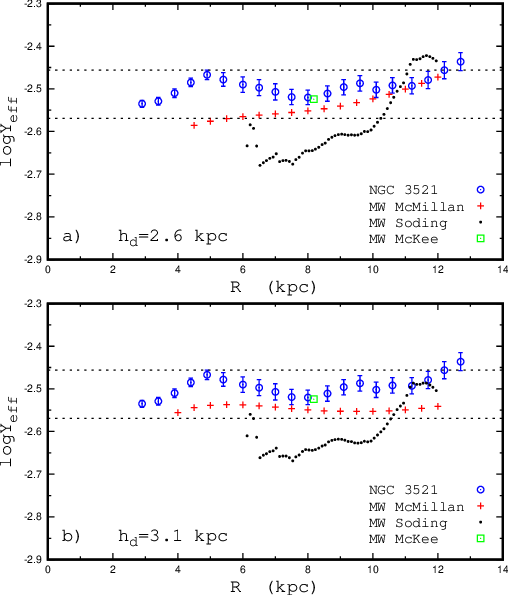}}
\caption{Comparison between the effective oxygen yields in the NGC~3521 and the Milky Way for the radial oxygen abundance distributions shown in Fig.~\ref{figure:c-r-oh}
  and the gas mass fraction distributions displied in Fig.~\ref{figure:c-r-mu}. 
  {\sl Panel {\bf a}:} The circles denote the effective oxygen yields in NGC~3521 as a function of radius and the bars show their uncertainties. 
  The plus signs designate the radial distribution of the effective oxygen yields in the Milky Way based on the $\Sigma_{\star}$ with the stellar disc scale length
  of $h_{d}$ = 2.6 kpc and on the $\Sigma_{gas}$ from \citet{McMillan2017}; the points are the yields based on the $\Sigma_{gas}$ from \citet{Soding2024}. The square marks the effective oxygen
  yield in the solar vicinity based on the $\Sigma_{gas}$ from \citet{McKee2015}. The lines show the upper and lower values of the empirical estimations of the oxygen yield. 
  {\sl Panel {\bf b}:} Same as panel (a), but for the stellar disc scale length of $h_{d}$ = 3.1 kpc. 
}
\label{figure:c-r-yeff}
\end{figure}

The simple model of chemical evolution of galaxies predicts that the oxygen abundance of the interstellar matter of a galaxy is related to the gas mass fraction $\mu$ and the oxygen
yield Y$_{\rm O}$ by the   formula
\begin{equation}
Z_{\rm O} = Y_{\rm O}\ln \left( \frac{1}{\mu} \right)   
\label{equation:cbm}
\end{equation}
\citep[][and references therein]{Pagel2009,Matteucci2012}, where Z$_{\rm O}$ is the oxygen abundance expressed in units of oxygen mass fraction relative to mass of the intersellar matter.
This simple model is based on the following assumptions: (i) an instantaneous recycling approximation, (ii) the element yield is constant (independent of metallicity), (iii) the system evolves
as a closed box (no matter exchange between the system and its surroundings). Assumptions (i) and (ii) are well justified for the oxygen evolution since the oxygen is a primary element
(the production of the oxygen is independent of the stellar metallicity) and the production and the ejection of newly synthesised oxygen by massive stars takes place in a short time compared
to the timescale of galactic evolution.  The gas exchange between the system and its surroundings (infall of low-metallicity gas and/or gas loss through the galactic winds) results in the
decrease in the oxygen abundance for a given value of the gas mass fraction   compared to the closed-box model; that is, it mimics the lowering of the oxygen yield
\citep[e.g.][]{Edmunds1990,Pilyugin1994}. Therefore, the deviation of the effective yield estimated using the measured oxygen abundance $Z_{O}$ and gas mas fraction $\mu$ 
\begin{equation}
 Y_{\rm eff} = Z_{\rm O} /\ln \left( \frac{1}{\mu} \right)   
\label{equation:yeff}
\end{equation}
from the true oxygen yield $Y_{o}$ can be used as an indicator of gas exchange between the system and its surroundings \citep[e.g.][]{Vilchez2019,LaraLopez2019,Garduno2023}. 
Empirical estimates of the oxygen yield $Y_{\rm O}$ for the metallicity scale defined by the  H\,{\sc ii} regions with the $T_{e}$-based oxygen abundances (the abundances used in this
study are compatible with this metallicity scale) result in $Y_{\rm O} = 0.0027$ \citep{Pilyugin2004}, $Y_{\rm O} = 0.0032$ \citep{Bresolin2004}, and between $Y_{\rm O}$ = 0.0030 and
$Y_{\rm O}$ = 0.0035 depending on the amount of oxygen incorporated into the dust grains \citep{Pilyugin2007}. 

Since the measured oxygen abundances are expressed in units of the number of oxygen atoms relative to hydrogen, while Z$_{\rm O}$ in Eq.~(\ref{equation:yeff}) is in units of mass fraction, 
we adopt the following conversion equation for the oxygen  \citep{Garnett2002}:
\begin{equation}
Z_{\rm O} = 12 \frac{O}{H}  .
\label{equation:zo-oh}
\end{equation}
We note that the oxygen abundances determined using the emission line spectra are the gas-phase abundances. \citet{Peimbert2010} estimated the dust depletion of oxygen
in Galactic and extragalactic H\,{\sc ii} regions and found that the fraction of oxygen atoms embedded in dust grains is a function of the oxygen abundance, which is about
0.12~dex for the metal-rich H\,{\sc ii} regions. This depletion has to be considered when the total (gas + dust) oxygen abundance is derived.

We estimate the values of the effective oxygen yields for different galactocentric distances within the optical radii in  NGC~3521 and the Milky Way using the distributions
of the gas mass fraction and the oxygen abundances described above. The simple model breaks down at low gas mass fractions because the term ln(1/$\mu$) blows up. Therefore, we
do not estimate the effective oxygen yields for $\mu \la$ 0.1. 
Figure~\ref{figure:c-r-yeff} shows the obtained values of the effective oxygen yield as a function of radius for  NGC~3521 and the Milky Way.
The lines mark the lower and upper values of the empirical estimations of the oxygen yield, $Y_{\rm O}$ = 0.0027 and  $Y_{\rm O}$ = 0.0035.
The uncertainty of the obtained value of the effective oxygen yield at a given radius is estimated in the following way. A random noise  ($\alpha \times$ uncertainty)
is added to each coefficient in the O/H -- $R$ relation (Eq.~\ref{equation:oh}) and to the mass surface densities of the atomic gas, molecular gas, and stars (these values and their
uncertainties are from \citealt{Leroy2008}), where $\alpha$ is a random number from --1 to 1. The value of the effective oxygen yield is estimated with the disturbed coefficients
and its difference with the effective oxygen yield for undisturbed coefficients  is determined. The mean difference for 1000 simulations is considered as the uncertainty of the value of
the effective oxygen yield. The obtained uncertainties of the values of the effective oxygen yields in NGC~3521 are shown by the bars in Fig.~\ref{figure:c-r-yeff}.

Figure~\ref{figure:c-r-yeff} shows that the obtained values of the effective oxygen yield in NGC~3521 are within the band outlined by the upper and lower
values of the empirical estimations of the oxygen yield. The oxygen abundances in NGC~3521 are close to those predicted by the simple model for the chemical evolution of galaxies.
This suggests that the mass exchange with the surroundings (the infall of low-metallicity gas and/or gas loss through the galactic winds) plays a small role, if any, in the current
chemical evolution of  NGC~3521.    

The values of the $Y_{eff}$ in the Milky Way obtained with the radial distribution of the gas mass surface density from \citet{McMillan2017}  are also
between the upper and lower values of the empirical estimations of the oxygen yield. 
There is a systematic trend in  $Y_{eff}$ with radius in the Milky Way:  the $Y_{eff}$ slightly increases from the centre to the periphery in the case of
the stellar disc scale length of $h_{d}$ = 2.6 kpc (panel (a) in Fig.~\ref{figure:c-r-yeff}). 
The systematic trend in  $Y_{eff}$ with radius in the Milky Way disappears for the stellar disc scale length of $h_{d}$ = 3.1 kpc, panel (b) in Fig.~\ref{figure:c-r-yeff}. 
The values of the  $Y_{eff}$ in the Milky Way obtained with the radial distribution of the gas mass surface density from \citet{Soding2024} are above the lower value of
the empirical estimations of the oxygen yield near the optical radius, at radii larger than $\sim$10.4 kpc (10.2 or 10.6 kpc depending on the adopted value of the stellar disc scale length).
Moreover, the  $Y_{eff}$ values are above the upper value of the empirical estimations of the oxygen yield at radii larger than 11.1 kpc for the case of $h_{d}$ = 2.6 kpc (panel (a) in
Fig.~\ref{figure:c-r-yeff}). The  $Y_{eff}$ values are below the lower value of the empirical estimations of the oxygen yield at smaller radii.
The effective yields are not estimated in the inner part of the Milky Way (at radii smaller than 6.1 kpc) because of the small value of  gas mass fraction (the simple model breaks down at
low gas mass fractions) or because of the oxygen abundances are not available. 

It was noted above that the gas exchange between the system and the surroundings (infall of low-metallicity gas and/or gas loss through  galactic winds) mimics the lowering of
the oxygen yield. The values of the effective yields below the lower value of the empirical estimations of the oxygen yield obtained in the Milky Way at radii between $\sim$6
and $\sim$10.4 kpc for the radial distribution of the gas mass surface density from \citet{Soding2024} can be considered as evidence    that  mass exchange with the
surroundings plays a significant role in the current evolution of this region in the Milky Way. 
It is important to note that the obtained value of the $Y_{eff}$ in the solar vicinity (where the measurement of the gas mass fraction is the most accurate and,
consequently, the estimation of the $Y_{eff}$ is more reliable) is  $Y_{eff}$ =  0.0030; that is, the effective oxygen yield in the solar vicinity is in agreement
with the empirical estimations of the oxygen yield.

Thus, the obtained values of the $Y_{eff}$ in  NGC~3521 are close to the $Y_{O}$; that is, the oxygen abundances in NGC~3521 are close to those predicted by the simple model for
the chemical evolution of galaxies. 
The values of the  $Y_{eff}$ in the outer part of the Milky Way obtained with the radial distribution of the gas mass surface density from \citet{McMillan2017}  are also
close to the $Y_{O}$ value; that is, the oxygen abundances in the Milky Way are close to those predicted by the simple model for the chemical evolution of galaxies.
This is evidence  that the mass exchange with the surroundings plays a small role, if any,   in the current chemical evolution of the outer part of the Milky Way (similar to what
is observed in NGC~3521).    

The values of the  $Y_{eff}$ in the outer part of the Milky Way obtained with the radial distribution of the gas mass surface density from \citet{Soding2024}  are more or less 
close to the empirical estimation of the oxygen yield near the optical radius,  and are below the lower value of the empirical estimations of the oxygen yield  at radii between $\sim$6
and $\sim$10.4 kpc. This can be considered as evidence  that  mass exchange with the surroundings can play a significant role in the chemical evolution of the
 outer part of the Milky Way, from $\sim$6 to $\sim$10.4 kpc. 
To make a solid conclusion of the role played by the  mass exchange with the surroundings in the chemical evolution of the outer part of the Milky Way, it should be established
which distribution of the surface gas mass density,  \citet{McMillan2017} or  \citet{Soding2024}, is a more adequate description of the gas distribution in the Milky Way.
The obtained value of   $Y_{eff}$ in the solar vicinity (where the measurement of the gas mass fraction is the most accurate \citep{McKee2015} and,
consequently,  the estimation of the effective oxygen yield is most reliable) is  $Y_{eff}$ =  0.0030, that is, the $Y_{eff}$  in the solar vicinity is in agreement
with the $Y_{O}$.

\section{Conclusions}

The IFU spectroscopy measurements of the galaxy NGC~3521 (structural Milky Way analogue) were carried out within the Metal-THINGS survey; the blue and red spectra were
measured in the IFU fibres for three pointings. The radial distributions of the oxygen abundance, the gas mass fraction, and the effective oxygen yield in NGC~3521 were
compared to that of the Milky Way, with the  aim of examining the similarity (or disagreement) of their chemical evolutions.

We found that the oxygen abundance in the inner region of the disc of  NGC~3521 is at a nearly constant level and the oxygen abundance gradient is negative at larger radii.
The change in the nitrogen abundance with radius is similar to that for oxygen with the break in the N/H distribution at a smaller radius than the O/H distribution break,
but the difference between the break radii is within the uncertainties of these values.
The obtained radial distributions of the oxygen and nitrogen abundances in  NGC~3521 are similar to those in other spiral galaxies with LS  gradients \citep{Pilyugin2024}.

The oxygen abundances in the two H\,{\sc ii} regions nearest to the centre of the Milky Way (at galactocentric distances of 4-5 kpc) are close to the binned oxygen abundances in
NGC~3521 at the same galactocentric distances. The central oxygen abundance in the Milky Way cannot be established due to the lack of  relevant abundance measurements  in
H\,{\sc ii} regions near the centre. The oxygen abundances in the outer part of the Milky Way are lower than in the outer part of NGC~3521.

The oxygen abundance at the optical radius of  NGC~3521 is lower than in the majority of other structural Milky Way analogues. The difference between the abundances at the
optical radius of  NGC~3521 and the Milky Way is smaller than in the majority of sMWAs, but the difference is as large as 0.17 dex.

The gas mass fraction in the outer part of  NGC~3521 is lower than that in the Milky Way. 
The obtained values of the effective oxygen yield, $Y_{eff}$, in  NGC~3521 are close to the empirical estimation of the oxygen yield, $Y_{O}$; that is, the oxygen abundances in NGC~3521
are close to those predicted by the simple model for the chemical evolution of galaxies. This indicates that the mass exchange with its surroundings (the infall of low-metallicity gas and/or
gas loss through  galactic winds) plays a small role, if any,   in the current chemical evolution of  NGC~3521.

The values of the $Y_{eff}$ in the MW obtained with the radial distribution of   $\Sigma_{gas}$ from \citet{McMillan2017} are close to the value of $Y_{O}$. This indicates that the mass
exchange with its surroundings also plays a small role in the chemical evolution of the MW, similarly to NGC~3521. The values of   $Y_{eff}$ in the MW obtained with   $\Sigma_{gas}$
from \citet{Soding2024}  are more or less close to   $Y_{O}$ near the optical radius and are below   $Y_{O}$ at radii between $\sim$6 and $\sim$10.4 kpc. This suggests that the
mass exchange with its surroundings plays a significant role in the chemical evolution of the outer part of the MW, in contrast to that in  NGC~3521. To make a solid conclusion of the
role of the mass exchange with the surroundings in the chemical evolution of the MW, it should be established which distribution of  $\Sigma_{gas}$,   from \citet{McMillan2017}
or from \citet{Soding2024}, is a more adequate description of the gas distribution in the MW. The value of   $\Sigma_{gas}$ in the solar vicinity \citep{McKee2015} is close to azimuthally
averaged surface mass density of gas at the solar galactocentric distance in the distribution from \citet{McMillan2017}. 


\begin{acknowledgements} 
We are grateful to the referee for his/her constructive comments. \\
We thank Dr. K.~Grasha for providing us with the measurements (emission line fluxes and coordinates) of their sample of H\,{\sc ii} regions in NGC~3521.  \\ 
We thank L.~S{\"o}ding for providing us with the necessary numerical data
on the radial distribution of the surface hydrogen density in the Milky Way.     \\
LSP acknowledges support from the Research Council of Lithuania (LMTLT), grant no. P-LU-PAR-24-38. \\
MALL acknowledges support from the Spanish grant PID2021-123417OB-I00, and the Ramón y Cajal program funded by the Spanish Government (RYC2020-029354-I) \\
MEDR acknowledges support from PICT-2021-GRF-TI-00290 of Agencia I+D+i (Argentina). \\
We acknowledge the usage of the HyperLeda database (http://leda.univ-lyon1.fr). \\
\end{acknowledgements}


\begin{thebibliography}{}
  
\bibitem[Arellano-C{\'o}rdova et al.(2020)]{Arellano2020}
         Arellano-C{\'o}rdova K.Z., Esteban C., Garc{\'\i}a-Rojas J., M{\'e}ndez-Delgado J.E., 2020, MNRAS, 496, 1051 
         
\bibitem[Arellano-C{\'o}rdova et al.(2021)]{Arellano2021}
         Arellano-C{\'o}rdova K.Z., Esteban C., Garc{\'\i}a-Rojas J., M{\'e}ndez-Delgado J.E., 2021, MNRAS, 502, 225
 
 \bibitem[Asari et al. (2007)]{Asari2007}        
          Asari, N. V., Cid Fernandes, R., Stasi\'{n}ska, G., et al. 2007, MNRAS, 381, 263
         
\bibitem[Bacchini et al.(2019)]{Bacchini2019}
         Bacchini C., Fraternali F., Pezzulli G., et al.,  2019, A\&A, 632, A127 

\bibitem[Baldwin et al. (1981)]{Baldwin1981}
         Baldwin J.A., Phillips M.M., Terlevich R., 1981, PASP, 93, 5
  
\bibitem[Begeman(1989)]{Begeman1989} 
         Begeman K.G., 1989, A\&A, 223, 47 
         
\bibitem[Bland-Hawthorn \& Gerhard(2016)]{BlandHawthorn2016}
         Bland-Hawthorn J., Gerhard O., 2016, ARA\&A, 54, 529 
         
\bibitem[Boardman et al.(2020a)]{Boardman2020a}
         Boardman N., Zasowski G., Seth A., et al., 2020a, MNRAS, 491, 3672 
         
\bibitem[Bresolin et al.(1999)]{Bresolin1999} 
         Bresolin F., Kennicutt R.C., Garnett D.R., 1999, ApJ, 510, 104 

\bibitem[Bresolin et al.(2004)]{Bresolin2004} 
         Bresolin F., Garnett D.R., Kennicutt R.C., 2004, ApJ, 615, 228

\bibitem[Bruzual \& Charlot (2003)]{BruzualCharlot2003}
         Bruzual, G., \& Charlot, S. 2003, MNRAS, 344, 1000

\bibitem[Campbell et al.(1986)]{Campbell1986}  
         Campbell A., Terlevich R., Melnick J., 1986, MNRAS, 223, 811 

\bibitem[Cardelli et al. (1989)]{Cardelli1989}
         Cardelli J.A., Clayton G.C., Mathis J.S., 1989, ApJ, 345, 245 

\bibitem[Cid Fernandes et al. (2005)]{CidFernandes2005}              
         Cid Fernandes, R., Mateus, A., Sodre, L., Stasi\'{n}ska, G., \& Gomes, J. M. 2005, MNRAS, 358, 363
                       
\bibitem[Cid Fernandes et al. (2010)]{CidFernandes2010}
         Cid Fernandes R.,  Stasi\'{n}ska G., Schlickmann M.S., et al., 2010, MNRAS, 403, 1036
   
\bibitem[Cid Fernandes et al. (2011)]{CidFernandes2011}
         Cid Fernandes R.,  Stasi\'{n}ska G., Mateus A., Vale Asari N., 2011, MNRAS, 413, 1687 

\bibitem[Croxall et al. (2016)]{Croxall2016}
         Croxall K.V., Pogge R.W., Berg D.A., Skillman E.D., Moustakas J., 2016, ApJ, 830, 4
          
\bibitem [D'Agostino et al. (2019)]{D'Agostino2019} 
          D'Agostino J.J., Kewley L.J., Groves B.A., et al., 2019, MNRAS, 485, L38
                   
\bibitem[Daigle et al.(2006)]{Daigle2006}
         Daigle O., Carignan C., Amram P., et al.,  2006, MNRAS, 367, 469 

\bibitem[Dame(1993)]{Dame1993}
         Dame T.M., 1993, in Holt S.S., Verter F., eds, American Institute of Physics Conference Series Vol. 278, Back to the Galaxy. pp 267–278,

\bibitem[Davis et al.(2014)]{Davis2014}
         Davis B.L., Berrier J.C., Johns L., et al., 2014, ApJ, 789, 124
          
\bibitem[de Blok et al.(2008)]{deBlok2008}
         de Blok W.J.G., Walter F., Brinks E., et al., 2008, AJ, 136, 2648
         
\bibitem[Deharveng et al.(2000)]{Deharveng2000}
         Deharveng L., Pe{\~n}a M., Caplan J., Costero R., 2000, MNRAS, 311, 329 
         
\bibitem[de Vaucouleurs \& Pence(1978)]{deVaucouleurs1978}
         de Vaucouleurs G., Pence W.D., 1978, AJ, 83, 1163 
          
\bibitem[de Vaucouleurs et al.(1991)]{RC3}
         de Vaucouleurs G., de Vaucouleurs A., Corvin H.G., Buta R.J., Paturel J., Fouque P. 1991, 
         Third Reference Catalog of bright Galaxies, New York: Springer Verlag (RC3)

\bibitem[Edmunds(1990)]{Edmunds1990} 
         Edmunds M.G., 1990, MNRAS, 246, 678
         
\bibitem[Fern{\'a}ndez-Mart{\'\i}n et al.(2017)]{Fernandez2017}  
         Fern{\'a}ndez-Mart{\'\i}n A., P{\'e}rez-Montero E., V{\'\i}lchez J.M., Mampaso A., 2017, A\&A, 597, A84 
         
\bibitem[Fielder et al.(2021)]{Fielder2021}  
         Fielder C., Newman J.A., Andrews B,H., et al., 2021, MNRAS, 508, 4459

\bibitem[Flynn et al.(2006)]{Flynn2006}
         Flynn C., Holmberg J., Portinari L., Fuchs B., Jahrei{\ss} H., 2006, MNRAS, 372, 1149 

\bibitem[Fraser-McKelvie et al.(2019)]{FraserMcKelvie2019}
         Fraser-McKelvie A., Merrifield M., Arag{\'o}n-Salamanca A., 2019, MNRAS, 489, 5030 

\bibitem[Gardu{\~n}o et al.(2023)]{Garduno2023}
         Gardu{\~n}o L.E., Zaragoza-Cardiel J., Lara-L{\'o}pez M.A., et al., 2023, MNRAS, 526, 2479 

\bibitem[Garnett (1992)]{Garnett1992} 
         Garnett D.R., 1992, AJ, 103, 1330

\bibitem[Garnett(2002)]{Garnett2002}
         Garnett D.R., 2002, ApJ, 581, 1019
         
\bibitem[Grasha et al.(2022)]{Grasha2022}
         Grasha K., Chen Q.H., Battisti A.J., et al., 2022, ApJ, 929, 118 

\bibitem[Gravity Collaboration(2019)]{Gravity2019}
         Gravity Collaboration, Abuter R., Amorim A., Baub{\'o}ck M., et al., 2019, A\&A, 625, 10 

\bibitem[Johnston et al.(2023)]{Johnston2023}
        Johnston V.D., Medling A.M., Groves B., et al., 2023, ApJ, 954, 77 

\bibitem[Hammer et al.(2007)]{Hammer2007}
         Hammer F., Puech M., Chemin L., Flores H., Lehnert M.D., 2007, ApJ, 662, 322 

\bibitem[Hill et al.(2008)]{Hill2008}
         Hill, G. J., MacQueen, P. J., Smith, M. P., et al. 2008, Proc. SPIE, 7014, 701470

\bibitem[Kalberla \& Dedes(2008)]{Kalberla2008}
         Kalberla P.M.W., Dedes L., 2008, A\&A, 487, 951 

\bibitem[Kauffmann et al.(2003)]{Kauffmann2003}
         Kauffmann G., Heckman T.M., Tremonti C., et al. 2003, MNRAS, 346, 1055
          
\bibitem[Kennicutt \& Garnett(1996)]{Kennicutt1996}
         Kennicutt R.C., Garnett D.R., 1996, 456, 504
          
\bibitem[Kennicutt at al.(2003)]{Kennicutt2003}
         Kennicutt R.C., Bresolin F., Garnett D.R., 2003, ApJ, 591, 801
                    
\bibitem[Kewley et al.(2001)]{Kewley2001}
         Kewley L.J., Dopita M.A., Sutherland R.S., Heisler C.A., Trevena J.  2001 ApJ, 556, 121
                 
\bibitem[Kewley \& Dopita (2002)]{Kewley2002}
         Kewley L.J., Dopita M.A., 2002, ApJS, 142, 35 
                   
\bibitem[Kreckel et al. (2019)]{Kreckel2019}
         Kreckel K., Ho I.-T., Blanc G.A., et al., 2019, ApJ, 887, 80 
         
\bibitem[Kubryk et al(2015)]{Kubryk2015}
         Kubryk M., Prantzos N., Athanassoula E., 2015, A\&A, 580, A126

\bibitem[Lacerda et al.(2018)]{Lacerda2018}
         Lacerda E.A.D., Cid Fernandes R., Couto G.S., et al., 2018, MNRAS, 474, 3727
          
\bibitem[Lang et al.(2020)]{Lang2020} 
         Lang P., Meidt S.E., Rosolowsky E., et al., 2020, ApJ, 897, 122 

\bibitem[Lara-L{\'o}pez et al.(2019)]{LaraLopez2019} 
         Lara-L{\'o}pez M.A., De Rossi M.E., Pilyugin L.S., et al., 2019, MNRAS, 490, 868 

\bibitem[Lara-L{\'o}pez et al.(2021)]{LaraLopez2021} 
         Lara-López M.A., Zinchenko I.A., Pilyugin L.S., et al. 2021, ApJ, 906, 42

\bibitem[Lara-L{\'o}pez et al.(2023)]{LaraLopez2023}
         Lara-L{\'o}pez M.A., Pilyugin L.S., Zaragoza-Cardiel J.,  et al., 2023, A\&A, 669, A25 
          
\bibitem[Leroy et al.(2008)]{Leroy2008}
         Leroy A.K., Walter F., Brinks E., et al., 2008, AJ, 136, 2782
                  
\bibitem[Leroy et al.(2021)]{Leroy2021} 
         Leroy A.K., Schinnerer E., Hughes A., et al., 2020, ApJS, 257, 43 
      
\bibitem[Licquia et al.(2015)]{Licquia2015b}
         Licquia T.C., Newman J.A., Brinchmann J., 2015, ApJ, 809, 96 
       
\bibitem[Licquia et al.(2016)]{Licquia2016b}
         Licquia T.C., Newman J.A., Bershady M.A., 2016, ApJ, 833, 220
         
\bibitem[L{\'o}pez et al.(2017)]{Lopez2017} 
         L{\'o}pez K.M., Heida M., Jonker P.G., et al., 2017, MNRAS, 469, 671 
      
\bibitem[L{\'o}pez et al.(2019)]{Lopez2019} 
         L{\'o}pez K.M., Jonker P.G., Heida M., et al., 2019, MNRAS, 489, 1249
      
\bibitem[Maiolino \& Mannucci(2019)]{Maiolino2019} 
         Maiolino R., Mannucci F., 2019, A\&A Rev., 27. 3 
         
\bibitem[Makarov et al.(2014)]{Makarov2014} 
         Makarov D., Prugniel P., Terekhova N., Courtois H., Vauglin I.. 2014, A\&A, 570, A13
         
\bibitem[Marasco et al.(2017)]{Marasco2017}
         Marasco A., Fraternali F., van der Hulst J.-M., Oosterloo T., 2017, A\&A, 607, 106 
         
\bibitem[Marino et al.(2013)]{Marino2013}
         Marino R.A., Rosales-Ortega F.F., S\'{a}nchez S.F., et al., 2013, A\&A, 559, A114     

\bibitem[Matteucci(2012)]{Matteucci2012}
         Matteucci F.,  2012, Chemical Evolution of Galaxies. Astronomy and Astrophysics Library, Springer-Verlag Berlin Heidelberg,  Germany 0

\bibitem[Matteucci \& Francois(1989)]{Matteucci1989}
         Matteucci F., Francois P., 1989, MNRAS, 239, 885

\bibitem[McGaugh(2016)]{McGaugh2016}
         McGaugh S.S., 2016, ApJ, 816, 42 
        
\bibitem[McKee et al.(2015)]{McKee2015}
         McKee C.F., Parravano A., Hollenbach D.J., 2015, ApJ, 814, 13 

\bibitem[McMillan(2017)]{McMillan2017}
         McMillan P.J., 2017, MNRAS, 465, 76

\bibitem[Mertsch \& Phan(2023)]{MertschPhan2023} 
         Mertsch P., Phan V.H.M., 2023, A\&A, 671, A54 
         
\bibitem[Mutch et al.(2011)]{Mutch2011}
         Mutch S.J., Croton D.J., Poole G.B., 2011, ApJ, 736, 84 
        
\bibitem[Nakanishi \& Sofue(2016)]{Nakanishi2016}
         Nakanishi H., Sofue Y., 2016, PASJ, 68, 5 

\bibitem[Oh et al.(2018)]{Oh2018} 
         Oh S.-H., Staveley-Smith L., Spekkens K., Kamphuis P., Koribalski B.S., 2018, MNRAS, 473, 3256 

\bibitem[Osterbrock \& Ferland(2006)]{Osterbrock2006}
         Osterbrock D.E., Ferland G.J., 2006, Astrophysics of Gaseous Nebulae and Active Galactic Nuclei.
         University Sciences- Books, Mill Valey, CA
    
\bibitem[Pagel(2009)]{Pagel2009}
         Pagel B.E.J., 2009, Nucleosynthesis and Chemical Evolution of Galaxies.  Cambridge University Press, Cambridge, UK.

\bibitem[Peimbert \& Peimbert(2010)]{Peimbert2010}
         Peimbert A., Peimbert M., 2010, ApJ, 724, 791

\bibitem[Pettini \& Pagel(2004)]{Pettini2004} 
         Pettini M., Pagel B.E.J., 2014, MNRAS, 348, L59
  
\bibitem[Pilyugin (1994)]{Pilyugin1994} 
         Pilyugin L.S., 1994, A\&A, 287, 387 

\bibitem[Pilyugin(2001)]{Pilyugin2001}
         Pilyugin L.S., 2001, A\&A, 373, 56
                    
\bibitem[Pilyugin et al.(2004)]{Pilyugin2004} 
         Pilyugin L.S., V\'{\i}lchez J.M., Contini T., 2004, A\&A, 425, 849 

\bibitem[Pilyugin et al.(2007)]{Pilyugin2007} 
         Pilyugin L.S., Thuan T.X., V\'{\i}lchez J.M., 2007, MNRAS, 376, 353 

\bibitem[Pilyugin et al.(2012)]{Pilyugin2012} 
         Pilyugin L.S., Grebel E.K., Mattsson L., 2012, MNRAS, 424, 2316 
        
\bibitem[Pilyugin \& Grebel(2016)]{Pilyugin2016} 
         Pilyugin L.S., Grebel E.K., 2016, MNRAS, 457, 3678 

\bibitem[Pilyugin et al.(2019)]{Pilyugin2019} 
         Pilyugin L.S., Grebel E.K., Zinchenko I.A., Nefedyev Y.A., V{\'\i}lchez J.M., 2019, A\&A, 623, A122 

\bibitem[Pilyugin et al.(2020)]{Pilyugin2020} 
         Pilyugin L.S., Grebel E.K., Zinchenko I.A., et al., 2020, A\&A, 639, A96 

\bibitem[Pilyugin et al.(2021)]{Pilyugin2021} 
         Pilyugin L.S., Cedr{\'e}s B., Zinchenko I.A., et al., 2021, A\&A, 653, A11

\bibitem[Pilyugin et al.(2022)]{Pilyugin2022} 
         Pilyugin L.S.,  Lara-L{\'o}pez M.A., V{\'\i}lchez J.M., et al., 2022, A\&A, 668, A5 

\bibitem[Pilyugin et al.(2023)]{Pilyugin2023} 
         Pilyugin L.S., Tautvai\~{s}ien\.{e} G., Lara-L\'{o}pez M.A., 2023, A\&A, 676, A57 

\bibitem[Pilyugin \& Tautvai\~{s}ien\.{e}(2024)]{Pilyugin2024} 
         Pilyugin L.S., Tautvai\~{s}ien\.{e} G., 2024, A\&A, 682, A41
          
\bibitem[Rudolph et al.(2006)]{Rudolph2006} 
         Rudolph A.L., Fich M., Bell G.R., et al., 2006, ApJS, 162, 346 
          
\bibitem[S{\'a}nchez et al.(2014)]{Sanchez2014} 
         S{\'a}nchez S.F., Rosales-Ortega F.F., Iglesias-P{\'a}ramo J., et al., 2014, A\&A, 563, A49

\bibitem[S{\'a}nchez et al.(2021)]{Sanchez2021} 
         S{\'a}nchez S.F., Walcher C.J., Lopez-Cob{\'a} C., et al., 2021, Rev. Mex. Astron. Astrofis., 57, 3 
                  
\bibitem[S{\'a}nchez et al.(2024)]{Sanchez2024} 
         S{\'a}nchez S.F., Lugo-Aranda A.Z., S{\'a}nchez Almeida J., et al., 2024, A\&A, 682, A71 

\bibitem[Schaefer et al.(2020)]{Schaefer2020}  
         Schaefer A.L., Tremonti C., Belfiore F., et al.,  2020, ApJL, 890, L3

\bibitem[Shaver et al.(1983)]{Shaver1983}  
         Shaver P.A., McGee R.X., Newton L.M., Danks A.C., Pottasch S.R., 1983, MNRAS, 204, 53 

\bibitem[Skrutskie et al.(2006)]{Skrutskie2006}
         Skrutskie M.F.. Cutri R.M., Stiening R., et al., 2006, AJ, 131, 1163

\bibitem[S{\"o}ding et al.(2024)]{Soding2024}
         S{\"o}ding L., Edenhofer G., En{\ss}lin T.A. et al., 2024, astro-ph:2407.02859

\bibitem[Storey \&  Zeippen (2000)]{Storey2000}
         Storey P.J., Zeippen C.J. 2000, MNRAS, 312, 813

\bibitem[V{\'\i}lchez et al.(2019)]{Vilchez2019} 
         V{\'\i}lchez J.M., Rela{\~n}o M., Kennicutt R., et al., 2019, MNRAS, 483, 4968 

\bibitem[Walter et al.(2008)]{Walter2008} 
         Walter F., Brinks E., de Blok W.J.G., et al., 2008, AJ, 136, 2563 

\bibitem[Warner et al.(1973)]{Warner1973} 
         Warner P.J., Wright M.C.H., Baldwin J.E., 1973, MNRAS, 163, 163

\bibitem[Zaritsky et al.(1994)]{Zaritsky1994}
         Zaritsky D., Kennicutt R.C., Huchra J.P., 1994, ApJ, 420, 87 

\bibitem[Zinchenko et al.(2016)]{Zinchenko2016}        
         Zinchenko I.A., Pilyugin L.S., Grebel E.K., Sánchez, S.F., Vílchez J.M., 2016, MNRAS, 462, 2715
         
\bibitem[Zinchenko et al.(2019)]{Zinchenko2019}
         Zinchenko I.A., Pilyugin L.S., Sakhibov F., et al., 2019, A\&A, 628, A55 

\end{thebibliography}
\end{document}